\documentclass[paper,11pt]{article}
\pdfoutput=1
\usepackage{jheppub}
\usepackage[T1]{fontenc}
\usepackage[utf8]{inputenc}
\usepackage[english]{babel}
\usepackage[babel]{csquotes}
\usepackage{amsmath}
\usepackage{amsfonts}
\usepackage{amssymb}
\usepackage{mathtools}
\usepackage{mathrsfs}
\usepackage{bm}
\usepackage{bbold}
\usepackage{braket}
\usepackage{graphicx} 
\usepackage{bmpsize}
\usepackage{comment}
\usepackage{booktabs}
\usepackage{array}

\newcommand{\al}{\alpha}

\newcommand{\ga}{\gamma}
\newcommand{\de}{\delta}
\newcommand{\D}{\Delta}
\newcommand{\ep}{\epsilon}
\newcommand{\varep}{\varepsilon}

\renewcommand{\th}{\theta}

\newcommand{\la}{\lambda}

\newcommand{\m}{\mu}
\newcommand{\n}{\nu}

\newcommand{\si}{\sigma}

\newcommand{\vp}{\varphi}
\newcommand{\p}{\phi}

\newcommand{\pa}{\partial}
\DeclarePairedDelimiter{\abs}{\lvert}{\rvert}

\newcommand{\beq}{\begin{equation}}
\newcommand{\eeq}{\end{equation}}
\newcommand{\mc}{\mathcal}
\newcommand{\mb}{\mathbb}
\newcommand{\mycomment}[1]{}
\newcommand{\wh}[1]{\widehat{#1}}
\newcommand{\fk}{f_\textup{bulk}}
\newcommand{\fy}{f_\textup{bdy}}
\newcommand{\yi}{y_i}
\newcommand{\yl}{y_L}
\newcommand{\yr}{y_R}

\preprint{HU-EP-15/08}

\title{ Boundary and Interface CFTs from the  Conformal Bootstrap}

\author[1]{Ferdinando Gliozzi,}
\author[2]{Pedro Liendo,}
\author[3]{Marco Meineri,}
\author[4]{Antonio Rago}

\affiliation[1]{Dipartimento di Fisica, Universit\`a di Torino, 
and Istituto Nazionale di Fisica Nucleare - sezione di Torino, Via P. Giuria 1 I-10125 Torino, Italy}
\affiliation[2]{IMIP, Humboldt-Universit\"at zu Berlin, IRIS Adelershof, Zum Gro\ss en Windkanal 6, 12489 Berlin, Germany}
\affiliation[3]{Scuola Normale Superiore, Piazza dei Cavalieri 7 I-56126 Pisa, Italy
and Istituto Nazionale di Fisica Nucleare - sezione di Pisa}
\affiliation[4]{Centre form Mathematical Science, 
 Plymouth University, Drake Circus, Plymouth PL4 8AA, United Kingdom}

\bigskip
\abstract{
We explore some consequences of the crossing symmetry for defect conformal field theories, 
focusing on codimension one defects like flat boundaries or interfaces.
We study surface transitions of the 3d Ising and other $O(N)$ models through numerical solutions 
to the crossing equations with the method of determinants.
In the extraordinary transition, where the low-lying spectrum of the surface operators 
is known, we use the bootstrap equations to obtain information on the bulk spectrum of 
the theory. In the ordinary transition the knowledge of the low-lying bulk spectrum allows 
to calculate the scale dimension of the relevant surface operator, which compares well with 
known results of two-loop calculations in 3d. Estimates of various OPE
coefficients are also obtained. 
We also analyze in 4-$\epsilon$ dimensions the renormalization group interface between the $O(N)$ model and 
the free 
theory and check numerically the results in 3d.}

\keywords{Conformal Field Theory, Conformal Bootstrap, Critical Ising Model}

\begin{document}

\maketitle

\section{Introduction and motivations.}

Conformal field theories (CFTs) play in many senses a preeminent role among quantum and statistical field theories. 
Such a privileged position is first granted by the flow of the renormalization group, whose fixed points are scale 
invariant theories, which usually show full conformal invariance \cite{Polchinski:1987dy, Dymarsky:2013pqa}.
 More generally, approximate scale invariance is a feature of systems in which a wide separation of scales makes 
the flow very slow in intermediate regions. Through the renormalization group, nature realizes the theories 
possessing the maximum amount of bosonic symmetry, both in condensed matter and in particle physics, in appropriate 
UV and IR regimes. Reversing the argument, one can understand a generic quantum field theory as a CFT deformed by a
 set of relevant operators. All perturbative analyses are in fact justified by the small size of relevant couplings in the UV limit. 
One can even pursue non-perturbative explorations of RG flows using the 
ultraviolet data as 
the only input \cite{Yurov:1989yu} (see also \cite{Hogervorst:2014rta, Coser:2014lla} and references therein). As 
a consequence, the importance of conformal invariance exceeds the experimental interest: conformal field theories
 are among the main actors in formal investigations of the space of quantum field theories, which has seen a huge
 development in recent times. Furthermore, they are an invaluable tool for studying quantum gravity, through the 
AdS/CFT correspondence \cite{Maldacena:1997re}.

The most striking feature of a generic CFT is that, however strongly coupled, it is completely described by 
two sets of numbers: the spectrum of scale dimensions of operators of every spin, and the Operator Product 
Expansion coefficients. This simplification occurs because the predictive power of the OPE is boosted by the conformal 
symmetry. On one hand, irreducible representations of the conformal group gather infinitely many operators, and 
the contribution to the OPE of every conformal family is labeled by the dimension and spin of the highest weight and is
 fixed up to a single coefficient. On the other hand, the OPE converges inside correlation functions 
\cite{Mack:1976pa}, and can be repeatedly used to reduce all of them to a sum over functions of the kinematic 
variables, the so called conformal blocks, one for each conformal family. This pairwise reduction can be carried 
out fusing operators in various different orders, so that sums over different blocks need to be equal. 
The crossing equations obtained this way provide constraints on the possible CFT data \cite{Ferrara:1973yt}, and 
after the seminal paper \cite{Rattazzi:2008pe}, a wealth of new results on the space of conformal field theories 
in dimensions greater than two were found by exploiting these constraints 
\cite{Rychkov:2009ij, Rattazzi:2010gj,Poland:2010wg,ElShowk:2012ht,Liendo:2012hy,Pappadopulo:2012jk,ElShowk:2012hu,
Gliozzi:2013ysa, Gaiotto:2013nva, El-Showk:2014dwa,Gliozzi:2014jsa,Beem:2013qxa,
Nakayama:2014yia,Chester:2014fya,Kos:2014bka,Chester:2014gqa,Beem:2014zpa,
Simmons-Duffin:2015qma,Bobev:2015vsa}. The method proposed in \cite{Poland:2010wg,ElShowk:2012hu} which we refer 
to as the \emph{linear functional method}, relies on unitarity to find forbidden regions in the space of the 
CFT data, by considering 
particular channels in the conformal block decomposition of a four-point function. At the boundary of these 
regions a spectrum which is crossing symmetric up to some maximum scale dimension can be extracted numerically 
\cite{ElShowk:2012hu}. It is not difficult to show that the four-point functions of local operators on the 
vacuum encode all of the constraints coming from crossing symmetry: however, one needs in principle all of them, 
and therefore
 the trial spectrum extracted from a specific correlator is not guaranteed to correspond to a unitary CFT. 
Sometimes it does, though \cite{ElShowk:2012ht}, or maybe a set of minimal hypotheses on the spectrum can be put 
in place to lower the bound disregarding uninteresting solutions which stand in the way \cite{Gaiotto:2013nva}. 
Another possibility is to consider more than one four-point function, so that further requirements on the spectrum 
can be made: for instance, internal symmetries differentiate the set of primaries appearing in different OPEs. 
This strategy was applied to the $3d$ Ising model in \cite{Kos:2014bka}, providing strong evidence that the 
presence of $\mathbb{Z}_2$ symmetry and two relevant primaries defines only one theory. 

The reader is referred to the aforementioned papers for a detailed explanation of the linear functional method. 
Here we shall employ a different technique, introduced in \cite{Gliozzi:2013ysa}, which we review in 
section \ref{sec:method}. The \emph{method of determinants} is based on the choice of a truncation of the 
spectrum, and directly provides an approximate solution to the crossing equation. It is independent from unitarity 
and can be applied to any correlator. On the other hand, it is not yet completely automated, and this makes it 
difficult in practice to deal with truncations involving many primaries. As a consequence, estimating the 
size of the systematic error is a delicate matter. We shall comment on this issue along the way. 

The aim of this paper is to apply the conformal bootstrap program to some examples of defect conformal field 
theories. These are theories in which the conformal group is broken down to the stabilizer of some hypersurface. 
We shall be concerned only with the case of a codimension one hyperplane, alias a flat interface, but the 
considerations in section \ref{sec:method} apply to generic flat conformal defects.
Motivations for studying conformal defects are again both phenomenological and abstract. For instance, conformal 
defects describe modifications of a $d$ dimensional QFT localized near a $p$ dimensional plane, with $p<d$, in 
the infrared limit, provided these modifications are not swept away by coarse graining, and scale invariance is 
enhanced to invariance under the conformal group $SO(p+1,1)$. The simplest example is of course a conformal boundary
 - that is, an interface between a non-trivial and the trivial CFT. Lower dimensional defects may correspond to 
magnetic-like impurities in a spin system, see for instance \cite{Billo:2013jda}, or to dispersionless 
fermions, acting as a source for the order parameter of some bosonic system \cite{Allais:2014fqa}, or to vortices 
in holographic superfluids and superconductors \cite{Dias:2013bwa}, etc. On the more abstract side, extended defects 
are probes of a system, and may be used to constrain properties of the bulk CFT. We shall in fact see this happening 
in the present study. Moreover, interfaces are a natural way to ``compare'' two theories, and may provide 
information on the geometric structure of the space of CFTs \cite{Douglas:2010ic}.

The conformal bootstrap was first applied to the boundary setup in \cite{Liendo:2012hy}, while the twist line 
defect defined in \cite{Billo:2013jda} was tackled in \cite{Gaiotto:2013nva}. Both papers are concerned with the 
$3d$ Ising model, and both used the linear functional method. In the latter, four-point functions of defect operator
s were considered, while the former focused on two-point functions of bulk operators. Correlators of defect 
operators are blind to bulk-to-defect couplings, but correlators of bulk primaries do not satisfy in general 
the positivity constraints required by the linear functional method, and \emph{ad hoc} assumptions were made in 
\cite{Liendo:2012hy}, motivated by computations in $2d$ and in $\ep$-expansion. Here we concentrate on the 
two-point function of bulk scalar primaries, using the method of determinants, which can be safely applied 
to this case. Since our main interest is again the $3d$ Ising model, we compare our results for the special 
and the extraordinary transitions with those of \cite{Liendo:2012hy}. We also find approximate solutions to 
the crossing equations corresponding to the ordinary transition, which cannot be studied with the linear 
functional. In the latter case we extended the analysis to the $O(N)$ models with $N=0,2,3$, where a 
comparison can be made with two-loop calculations. The main results are summarized in the tables 
\ref{tab:1} and \ref{tab:2}. 

In the end, we initiate the study of an example of RG domain wall, an interface between two CFTs connected by the
 renormalization group, which is obtained by turning on a relevant deformation on half of the space and flowing 
to the IR. Specifically, we study the flow triggered by the $(\p^2)^2$ coupling in a bosonic theory. We give a 
first order description in $\epsilon$-expansion which applies to models with $O(N)$ symmetry and can be easily 
generalized to other perturbation interfaces. We then focus on the Ising model when looking for a numerical 
solution to the crossing equations in $3d$.

The structure of the paper is as follows. In section \ref{sec:method} we review the general features of conformal
 field theories in the presence of defects, and we explain the method of determinants. Section \ref{sec:boundary}
 is devoted to the study of the boundary CFTs associated to the 3d Ising and other spin systems. We define 
and study the domain wall in section \ref{sec:interface}. Finally, we draw our conclusions in section 
\ref{sec:conclusions}.  Appendix \ref{sec:details} contains some details of 
the $\ep$-expansion computations.

\section{Defect CFTs and the method of determinants.}
\label{sec:method}

The constraints imposed by conformal symmetry on correlation functions near a boundary were analyzed in \cite{McAvity:1995zd} (see also \cite{McAvity:1993ue}), and the boundary bootstrap was set up in \cite{Liendo:2012hy}, from which we borrow the notation. Here we review the necessary material, and then introduce the method of determinants. A general $p$-dimensional defect differs from the codimension one case for the residual $SO(d-p)$ symmetry generated by rotations around the defect. This is just a flavor symmetry for the defect operators, but induces some differences when it comes to bulk-to-defect couplings. Although most of what we shall say applies to a generic flat defect, in this paper we shall be concerned with the codimension one case. Therefore, further reference to the general case are limited to some side comments.

Correlation functions of excitations living at the defect are the same as in an ordinary $(d-1)$-dimensional CFT, and are completely characterized by the spectrum of scale dimensions ($\widehat\D_l$) and the coefficients of three-point functions ($\,\widehat\la_{lmn}$). We shall later need one more piece of information. While no conserved stress-tensor is expected to exist on the defect, a protected scalar operator of dimension $d$ $-$ or $p+1$ in the general case $-$ is always present: the displacement operator, which we call $D(x^a)$, measures the breaking of translational invariance, and is defined by the Ward identity for the stress-tensor:
\beq
\pa_\m T^{\m d}(x)=-D(x^a)\,\de(x^d).
\label{eq:stressWard}
\eeq
Here we denoted by latin indices the directions along the defect, which is placed at $x^d=0$, while Greek letters run from $1$ to $d$. Similarly, for every bulk current whose conservation is violated by the defect, a protected defect operator exists.

In the bulk, there is of course the usual OPE. For scalar primaries,
\beq
O_1(x) O_2(y) = \frac{\de_{12}}{(x-y)^{2\D_1}} + \sum_k \la_{12k} C[x-y, \pa_y] O_k(y)\, ,
\label{eq:bulkOPEscalar}
\eeq 
where $C[x-y, \pa_y]$ are determined by conformal invariance, and we isolated the contribution of the identity. One can also fuse a local operator with the defect. The bulk operator is thus turned into a sum over defect primaries. The bulk-to-defect OPE for a scalar primary can be written
\beq
O_1(x) = \frac{a_1}{\abs{2x^d}^{\D_1}} + \sum_l \mu_{1l} D[x^d,\pa_a] \widehat O_l(x^b)\,,
\label{eq:bdyOPEscalar}
\eeq
where we denoted defect operators with a hat. Again, the differential operators $D[x^d,\pa_a]$ are fixed by conformal invariance. Similar OPEs can be written for bulk tensors. The $\la_{12k}$'s in eq. \eqref{eq:bulkOPEscalar} are the coefficients of three-point functions without the defect, while $\mu_l$ is the coefficient of the correlator $\braket{O(x)\widehat O_l(y^a)}$, otherwise fixed by conformal symmetry. Even if, for the sake of simplicity, some abuse of notation is present\footnote{For instance, the coefficient $\m_{\p^2D}$ in free theory appears in the two point function $\braket{\frac{\p^2}{\sqrt{2N}}D}$.}, in this paper all OPE coefficients refer to canonically normalized operators, with one exception: the normalization of the displacement operator is fixed by eq. \eqref{eq:stressWard}. Taking the expectation value of both sides in eq. \eqref{eq:bdyOPEscalar} one sees that a scalar acquires a one-point function proportional to $a_O$, the coefficient of the identity in the bulk-to-defect OPE. It is not difficult to prove that tensors do not acquire an expectation value in the presence of a codimension one defect. They do, instead, if they are even spin representations and the defect is lower dimensional.

Let us now derive the easiest crossing equation involving the OPEs \eqref{eq:bulkOPEscalar} and \eqref{eq:bdyOPEscalar}. Consider the two-point function $\braket{O_1(x)O_2(x')}$. One can decompose it into the bulk channel by plugging in eq. \eqref{eq:bulkOPEscalar}: a sum over one-point functions is obtained, that is, a sum over the coefficients $\la_{12k}a_k $ multiplying some known functions of the kinematic variables. Or, one can substitute both operators with their Defect OPE, and in this case the sum involves the quantities $\m_{1l}\m_{2l}$. In order to write explicitly the equality of the two conformal block decompositions, let us introduce the conformal invariant combination
\beq
\xi=\frac{(x-x')^2}{4x^d x'^d}.
\eeq
This cross-ratio is conveniently positive when both points are chosen in the half-plane $x^d>0$. This is not the case when considering bulk operators on opposite sides of an interface. Moreover, in this setup the bulk OPE is not defined. The issue is solved by folding the system and treating it as a boundary CFT: the folding trick provides us with a trivial OPE, fixed by the absence of local interactions between the two primaries. We shall have more to say on this point in section \ref{sec:interface}. For now, we just point out that the natural cross-ratio is the one constructed from a point and the mirror image of the second one, and it is again positive. We assume $\xi \geq0$ in the rest of this section.

Conformal symmetry justifies the following parametrization:
\beq
\braket{O_1(x) O_2(x')} = \frac{1}{(2x^d)^{\D_1}(2x'^d)^{\D_2}}\, \xi^{- (\D_1 + \D_2)/2} 
G_{12}(\xi).
\label{eq:twoptscalar}
\eeq
Then the crossing equation can be written as a double decomposition of the function $G_{12}(\xi)$:
\beq
G_{12}(\xi) = \de_{12} + \sum_{k} \la_{12k}\, a_k \,\fk(\D_{12},\D_k;\xi) =
\xi^{(\D_1+\D_2)/2} \left( a_1a_2 + \sum_{l} \mu_{1l}\,\mu_{2l} \,\fy(\widehat{\D}_l;\xi) \right)\,,
\label{eq:crossingsy}
\eeq
where \cite{McAvity:1995zd}
\begin{subequations}
\begin{align}
&\fk(\D_{12},\D,\xi) = \xi^{\D/2} \,
{}_2 F_1 \left( \frac{1}{2}(\D_1-\D_2+\D), \frac{1}{2}(\D_2-\D_1+\D); 
\D +1-\frac{d}{2},-\xi \right), \\
&\fy(\D,\xi) = \xi^{-\D} \,
{}_2F_1\left(\D,\D+1-\frac{d}{2};2\D+2-d;-\frac{1}{\xi}\right).
\end{align}
\label{eq:blocks}
\end{subequations}

It is worth noticing that  the conformal blocks of the boundary channel in $d=3$ can be expressed as elementary algebraic functions, namely,
\beq
\fy(\Delta,\xi)\vert_{d=3}=\frac1{2\sqrt{\xi}}
\left(\frac4{1+\xi}\right)^{\Delta-\frac12}\left[1+
\sqrt{\frac\xi{1+\xi}}\right]^{-2(\Delta-1)}\,.
\eeq
This is of course of great help in  numerical calculations.

 Before describing how to extract information from eq. \eqref{eq:crossingsy}, we make some side remarks. 
The set $\{\widehat\D_l,\widehat\la_{lmn} ,
\D_i,\la_{ijk}, a_i, \mu_l\}$ is in fact redundant: by repeatedly applying the bulk-to-defect OPE one can reduce all correlators to correlators of defect operators, therefore the $\la_{ijk}$ are in principle unnecessary to solve the theory. 
However, it is easy to realize that all crossing equations constraining the bulk-to-defect couplings $\m_l$ also involve the bulk three-point function coefficients. One is naturally led to the following question: what is the minimal set of correlators encoding all the crossing symmetry constraints of a Defect CFT? All the four-point functions of defect operators are surely in the number, the proof being the usual one (see for instance \cite{Rychkov:epfl}). A similar argument shows that all the other crossing equations of a generic correlator of bulk and defect primaries are automatically satisfied once the three-point functions $\braket{O_1O_2\,\widehat O}$ are crossing symmetric.
 In the rest of this paper we explore the case $\widehat O = \mathbb{1}$, leaving for future work the general case.

Let us now turn our attention back to eq. \eqref{eq:crossingsy} that we rewrite in the following form
\beq
-\sum_{k} \la_{12k}\, a_k \,\fk(\D_{12},\D_k;\xi) +
\xi^{(\D_1+\D_2)/2} \left( a_1 a_2 + \sum_{l} \mu_{1l} \,\mu_{2l} \,
\fy(\widehat{\D}_l;\xi) \right)=\de_{12}\,.
\label{eq:crosy}
\eeq
 In most situations, an infinite number of operators contributes to both channels, which makes the crossing constraint difficult to exploit. The strategy described in \cite{Gliozzi:2013ysa} can be summarized in the following way. First, we trade one functional equation for infinitely many linear 
equations: one for each coefficient of the Taylor expansion around, say, $\xi=1$. Then we truncate both the Taylor 
expansions, keeping only the first $M$ derivatives, and the spectrum, keeping the first $N$ operators in total from the 
two channels. The bulk identity is excluded from the count. We denote this truncation with a triple $(n_{bulk},n_{bdy},s)$, the three numbers counting respectively bulk and boundary operators of non vanishing dimension, and the presence ($s=1$) or absence ($s=0$) of the boundary identity. We obtain this way a finite system, at the price of introducing a systematic error, coming from the disregarded higher 
order derivatives and heavier operators:
%\begin{align}
%&-\sum_k^{n_{bulk}} p_k \left.\fk^k \right|_{\xi=1}+
%\sum_l^{n_{bdy}} q_l \left.\fy^l\right|_{\xi=1} +a_1a_2= \de_{12},& \hspace{-0.04\columnwidth} n_{bulk}+n_{bdy}+s=N& \notag \\
%&-\sum_k^{n_{bulk}} p_k \left.\pa_\xi^n\fk^k\right|_{\xi=1}+
%\sum_l^{n_{bdy}} q_l \left.\pa_\xi^n \left(\xi^{(\D_1+\D_2)/2}\fy^l\right)\right|_{\xi=1} +a_1a_2
%\left.\pa_\xi^n\xi^{(\D_1+\D_2)/2}\right|_{\xi=1}=0,& n=1,\dots, M,&
%\label{eq:linearsys}
%\end{align}
\begin{multline}
-\sum_k^{n_{bulk}} p_k \left.\fk^k \right|_{\xi=1}+
\sum_l^{n_{bdy}} q_l \left.\fy^l\right|_{\xi=1} +a_1a_2= \de_{12}, 
\qquad \qquad \ \  n_{bulk}+n_{bdy}+s=N~,  \\
-\sum_k^{n_{bulk}} p_k \left.\pa_\xi^n\fk^k\right|_{\xi=1}+
\sum_l^{n_{bdy}} q_l \left.\pa_\xi^n \left(\xi^{(\D_1+\D_2)/2}\fy^l\right)\right|_{\xi=1} +a_1a_2
\left.\pa_\xi^n\xi^{(\D_1+\D_2)/2}\right|_{\xi=1}=0, \\
n=1,\dots, M~,
\label{eq:linearsys}
\end{multline}
where we used a shorthand notation for the OPE coefficients 
$p_k=\la_{12k}a_k,\,q_l=\m_{1l}\m_{2l}\,$. 
Let us focus for definiteness on the case of two identical external scalars, 
$\de_{12}=1$. The $p_k$'s, $q_l$'s and $a_1^2$  are the unknowns 
of a linear system whose coefficients depend nonlinearly on the 
bulk and defect spectra. Choosing $M\ge N$, the homogeneous system, i.e. the second line in \eqref{eq:linearsys}, admits a non-trivial solution if 
and only if all the $\binom{M}{N}$ minors of the system vanish. This condition provides a set of non-linear equations in the $N$ unknown scale dimensions. When this set admits a (numerical) solution we say that the the two-point function under study is truncable. In such a case, inserting the obtained (approximate) spectrum in the complete linear system (\ref{eq:linearsys}), we get the OPE coefficients.

Notice that every consistent CFT data is in particular a solution to this crossing equation. Therefore, some input has to be provided: here we are implicitly assuming that the external dimensions are known, and in fact this is going to be the strategy when we try to isolate the $3d$ Ising model. One does not expect to find an exact solution for a generic truncation: heavier defect and bulk operators become more and more important when moving respectively towards the bulk ($\xi\to0$) or the defect ($\xi\to\infty$), therefore we expect a good truncation to require $N$ to grow with $M$. In practice, in this work we usually choose 
$M=N+1$, and we find that the space of solutions to the system of nonlinear equations has in general non-zero dimension. By fixing the free parameters with the best known values of the lowest lying bulk primaries, we give predictions for the low lying defect spectrum and for heavier primaries.

\begin{figure}[htb]
\centering
\includegraphics[scale=0.8]{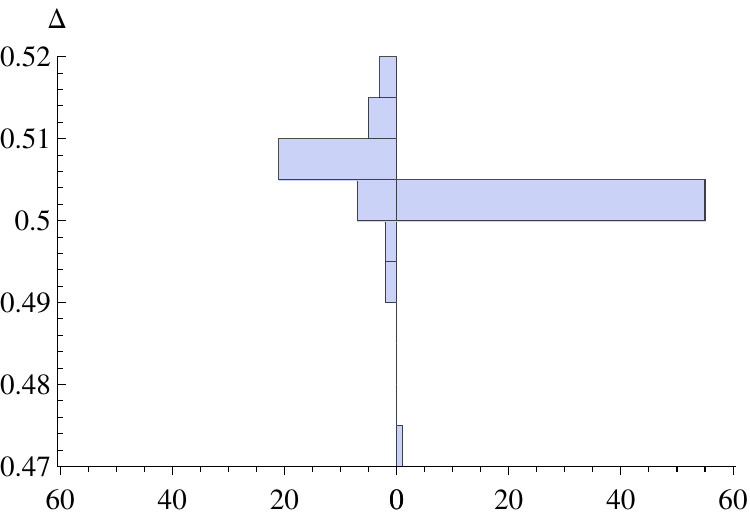}
\includegraphics[scale=0.8]{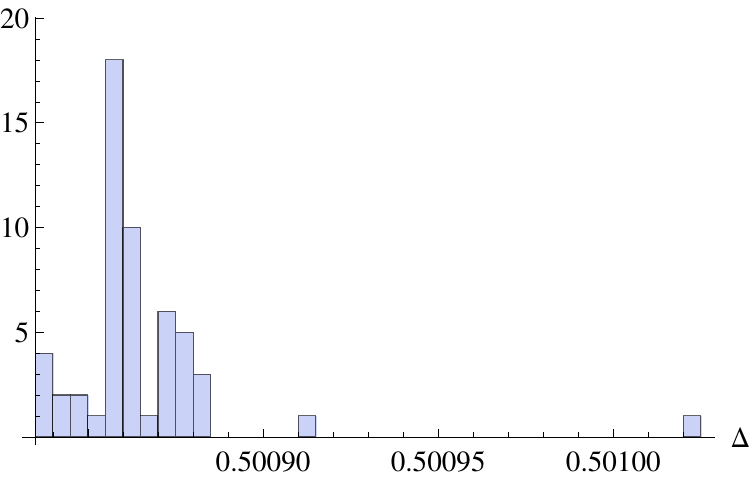}
\caption{Top panel: paired histograms of the solutions of two different truncations of the crossing equations for the ordinary transition of the 2d Ising model. 
Left: histogram for  the scale dimensions of 
the first boundary operator in the (2,1,0) truncation. The 
exact result is at $\widehat{\Delta}=\frac12$. Right: 
the corresponding histogram for the 
(4,3,0) truncation. Bottom panel: a more detailed view of the latter histogram.}
\label{fig:histo}
\end{figure}

\begin{figure}[htb]
\centering
\includegraphics[width=11  cm]{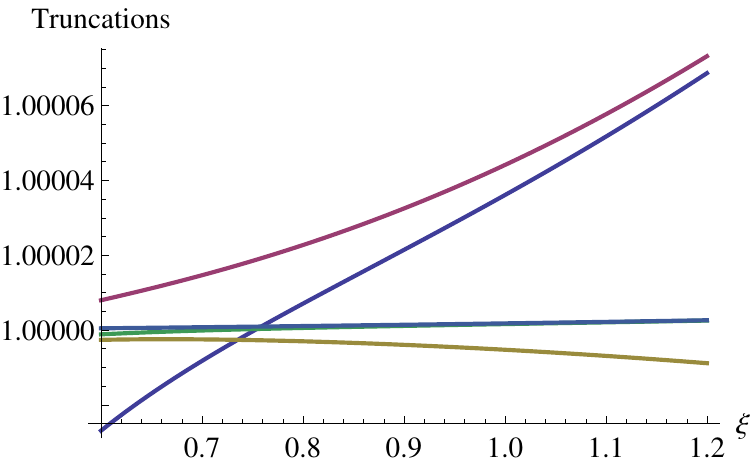}
\caption{The left-hand-side of the sum rule (\ref{eq:crosy}) for 
various truncations $(n_{bulk},n_{bdy},0)$ of the two-point function of the 2d Ising model in the ordinary transition. Only in the  $n_{bulk}\to\infty$, 
$n_{bdy}\to\infty$ limit the sum rule is saturated.}
\label{fig:truncations}
\end{figure}

As a general rule, a finite truncation of the crossing symmetry equations is a good approximation of a given CFT if the missing operators can be 
consistently put at $\Delta=\infty$ or at zero coupling. When a trial 
spectrum has been found, one can check its stability by adding one operator and one derivative. It turns out in most cases that the scaling dimension of the new operator acts as a free parameter which can vary in a fixed range. We use the solution for predictions only if it does not depend very strongly on this parameter. This gives a way of controlling the systematic error, albeit not an algorithmic one. Let us also observe that the general agreement with the results of the epsilon expansion suggests that the error is rather small, at least for what concerns the boundary case. Another important 
check comes from the Ward identity associated with the displacement operator, which, as we shall see, yields non-trivial relations among the CFT data. These relations are perfectly verified by the numerical solutions, as described in the next section.

Another parameter to be considered in order to check the quality of a given truncation is the spread of the solutions. As soon as the number $M$ of equations exceeds the number of unknowns, the system is over-determined and can be split in consistent subsystems, each of them giving in principle a different solution. The spread of these solution gives a rough estimate of the error. In the cases where the exact solution is known the narrower is the spread the closer is the solution to its exact value. This is the case for instance of the four-point function of the free scalar massless theory in any dimension \cite{Gliozzi:2013ysa}.
On the contrary  large spreads are associated to large systematic errors due 
to too rough approximations of the crossing equations. A clear illustration of this behavior can be found in the ordinary transition of the 2d Ising model, where the  exact two-point function is known \cite{DiFrancesco:1997nk}.
Assuming we already know the bulk spectrum, we can start considering the truncation (2,1,0) to evaluate the scale dimensions of the first surface operator. We have to look at the zeros of $3\times3$ determinants. Taking for instance 8 derivatives we have 56 equations whose solutions are plotted in the histogram of of fig.
\ref{fig:histo}. Their large spread is associated with a rather rough approximation of the sum rule (\ref{eq:crosy}) as fig. \ref{fig:truncations} shows. The same figure  points out also that the truncation (4,3,0) is much better. In this case the unknowns are the dimensions of the three surface operators. The consistent subsystems are made of sets of three $7\times7$ determinants. With 8 derivatives we have again 56 possible solutions. Their spread  is 
drastically reduced and the mean value is closer to the exact one, 
as fig. \ref{fig:histo} shows.  
We anticipate that all the solutions considered in the next 
section have a microscopic spread (see e.g. fig. \ref{fig:1} and fig. 
\ref{fig:3}).

\section{The boundary bootstrap and the $3d$ Ising and $O(N)$ models.}
\label{sec:boundary}
In this Section we shall consider the boundary conformal field theories (BCFTs) associated with the Ising model and other magnetic systems. Specifically, the IR properties of the surface transitions in these systems are controlled by RG fixed points, which of course are described by just as many Defect CFTs. We denote with $\sigma(x)$ the scalar field (i.e. the order parameter of the theory) and with $\widehat{\sigma}$ the corresponding surface operator.
The surface  Hamiltonian associated with a flat $d-1$ dimensional boundary of a semi-infinite system can be written in terms of the three relevant surface operators (see for instance \cite{Cardy:1996xt})
\beq
H=\int d^{d-1}x\left(c\widehat{\sigma}^2+h_1\widehat{\sigma}+h_2
\pa_z\widehat{\sigma}\right)\,.
\eeq
Here $z\equiv x^d$ is the coordinate orthogonal to the boundary. 
This Hamiltonian has three fixed points
\begin{align}
O:&\quad h_1=h_2=0,\,c=+\infty\,; \\
E:&\quad h_1=h_2=0,\,c=-\infty\,; \\
S:&\quad h_1=h_2=c=0\,.
\end{align}
Near the first fixed point the configurations with $\widehat{\sigma}\not=0$ are exponentially suppressed, then $\widehat{\sigma}=0$ (i.e. Dirichlet boundary condition). This fixed point controls the ordinary transition. The only relevant surface operator in this phase is $\pa_z\widehat{\sigma}$. The fixed point with $c=-\infty$ favors the configurations with $\widehat{\sigma}\not=0$: it is associated with the extraordinary transition, where the $\mb{Z}_2$ symmetry is 
broken and no relevant surface operator can couple with it; the lowest dimensional surface operator, besides the identity, is the displacement, whose scaling dimension is $d$. The fixed point with $c=0$ controls the special transition, a multicritical phase with two relevant primaries. The even operator 
$\widehat{\sigma}^2$ is responsible for the flow of $c$ to $\infty$ or $-\infty$ according to the initial sign, while the odd one, $\widehat{\sigma}$, is the symmetry breaking operator of this phase, characterized by the Neumann 
boundary condition $\pa_z\widehat{\sigma}=0$. We omitted a classically marginal coupling, $\pa_z \widehat{\si}^2$, because it vanishes with both Neumann and Dirichlet boundary conditions, and it cannot be turned on in the extraordinary transition, where there is no local odd relevant excitation. We shall come back to this operator when considering the RG domain wall.

One important question to address within a BCFT is how to find the 
scale dimensions of the surface operators and their OPE coefficients in terms of the bulk data. This problem has been completely solved in 2d \cite{Cardy:1991tv} thanks to the modular invariance. In $d>2$ useful information can be extracted by  the epsilon expansion and other perturbative methods. Recently the conformal bootstrap approach has been shown to be very promising \cite{Liendo:2012hy}. Here we face this problem with the method of determinants.

We study the 2-point function $\braket{\sigma(x)\sigma(y)}$.
The general criterion we use to classify the surface transition associated with a specific truncation $(n_{bulk},n_{bdy},s)$ of the crossing  symmetry
 equations (\ref{eq:linearsys}) is based on three steps. First, we verify that the solution is compatible with a unitary theory by requiring the positivity  of all the non-vanishing couplings $\mu_a^2$ $(a=1,2,\dots,n_{bdy})$. Then we look at the sign of the couplings to the bulk blocks $a_k\lambda_{\sigma\sigma k}$ $(k=1,\dots,n_{bulk})$. As in \cite{Liendo:2012hy}, we will assume that the
ordinary transition is signaled by the presence of at least one negative coupling in the bulk channel. On the other hand, positivity of the couplings indicates the extraordinary or the special transition, depending on the presence or absence of the surface identity. 
We should point out that these assumptions have not been proven. However, the results of this work seem to confirm them, serving as a consistency check on the whole setup.  

\subsection{The ordinary transition.}
\begin{figure}
\centering
\includegraphics[width=10 cm]{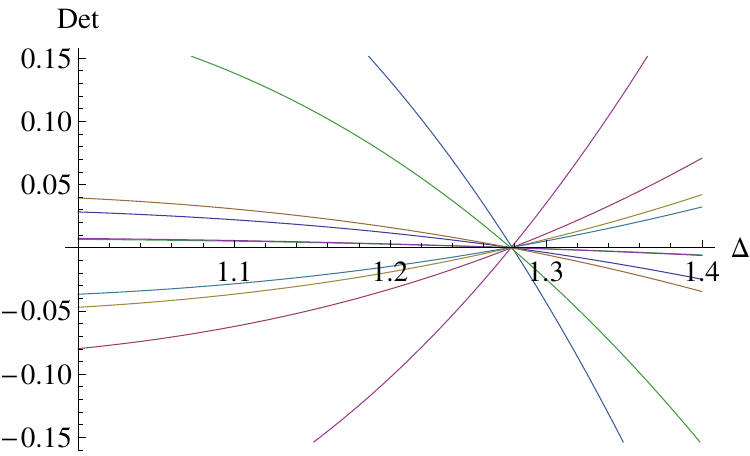}
\caption{Plot of the 10  $3\times3$ minors  made with the first 5 derivatives 
of the conformal blocks associated with $\varepsilon$, $\varepsilon'$ and $\widehat O$ as functions of $\D_{\widehat O}$. They all vanish approximately at he same point, selecting the allowed value of $\D_{\widehat O}$.}
\label{fig:1}
\end{figure}

We start by considering what is perhaps the simplest successful truncation 
of eq. \eqref{eq:linearsys}, corresponding to the fusion rules 
\begin{align}
&\sigma\times\sigma \sim 1+\varepsilon+\varepsilon', & &\textup{bulk channel,} \notag\\
&\sigma \sim\widehat{O},& & \textup{boundary channel.}
\end{align} 
This truncation 
is denoted by the triple (2,1,0). The system (\ref{eq:linearsys}) admits a solution if and only if the $3\times3$ determinants made with the derivatives of the conformal blocks associated with $\varepsilon,\varepsilon',\widehat{O}$ vanish. 
We assume that the scale dimensions of $\sigma,\varepsilon$ and $\varepsilon'$ are known
$(\Delta_\sigma=\frac12+\frac\eta2;\ 
\Delta_\varepsilon=3-1/\nu;\ \Delta_{\varepsilon'}=3+\omega$, see table \ref{tab:1}) and in this particular case the only unknown scale dimension is $\Delta_{\widehat{O}}$. Fig. \ref{fig:1} shows the values of few determinants of this kind. 
Clearly they all apparently vanish  at the same point. 
In fact there is a microscopic spread of the solutions and we find  
$\Delta_{\widehat{O}}=1.276(2)$.  The solution of the complete linear system 
yields a negative $a_{\varepsilon}\lambda_{\sigma\sigma \varepsilon}$, thus, according 
to the above criterion, we are faced with the ordinary transition of the 3d Ising model. Hence, $\widehat{O}$ has to be identified with $\pa_z\widehat\sigma$. A two-loop calculation in the 3d $\phi^4$ model yields \cite{Diehl:1998mh} 
$\Delta_{\pa_z\widehat\sigma}\simeq1.26$ in good agreement with our result.

\begin{table}[htb]
\centering
\begin{tabular}{*{4}{>{$}c<{$}}}
\toprule
N   &\eta              &\nu                  &\omega \\
\midrule
0   &  0.0314(32)   &    0.5874(2)    &   0.812(16) \\
1   & 0.03627(10)  &  0.63002(10)    &   0.832(6)  \\
2  &  0.0380(4)     &  0.67155(27)   & 0.789(11) \\
3  &   0.0364(6)    &   0.7112(5)    &   0.782(13) \\
\bottomrule
\end{tabular}
\vspace{0.025\textheight}
\begin{flushleft}
\begin{tabular}{*{4}{>{$}c<{$}}}
\toprule
\multicolumn{4}{c}{$\Delta_{\partial_z\widehat{\sigma}}$} \\
\midrule
N  &   \textup{2-loop}& \textup{Monte Carlo}&\textup{Bootstrap}  \\
0  &    1.33  &-             &   1.332(6)   \\
1  &     1.26 &1.2751(6)     &   1.276(2)  \\
2  &    1.211 &1.219(2)      &   1.2342(9)  \\
3  &    1.169 &1.187(2)     &    1.198(1)  \\
\bottomrule
\end{tabular} 
\end{flushleft}\hspace{0.16\columnwidth}
\begin{flushright}\vskip -4.1 cm\begin{tabular}{*{4}{>{$}c<{$}}}
\toprule
N   & a_\varep\la_{\si\si\varep}  &   a_{\varep'} \la_{\si\si\varep'}  & \mu_{\widehat \D}^2 \\
\midrule
0   &   -0.8447(34)  &        0.0366(17)   &   0.692(1)     \\
1   &   -0.789(3)    &        0.042(1)     &   0.755(13)    \\
2   &   -0.747(1)    &        0.0488(4)    &   0.80022(5)   \\
3   &   -0.710(1)    &        0.0509(6)    &   0.8395(6)    \\
\bottomrule
\end{tabular}
\end{flushright}
\caption{The first table collects the input parameters. The second one is a comparison between two-loop calculations \cite{Diehl:1998mh}, Monte Carlo simulations (reference \cite{Hasenbusch:2011} for $N=1$ and reference \cite{Deng:2005} for $N>1$) and our bootstrap results for the scaling dimension of the surface operator $\partial_z\widehat{\sigma}$ in the ordinary transition of 3d 
$O(N)$ models. The last three columns collect our results 
for the OPE coefficients. The critical 
indices $\eta$ and $\nu$ for 
$N=0,1,2,3$ are taken respectively from references \cite{Prellberg},
\cite{Hasenbusch:2011yya}, \cite{Campostrini:2000iw} and \cite{Campostrini:2002ky}. Those for $\omega$ from \cite {Guida:1998bx}. }
\label{tab:1}
\end{table}

This solution admits  a straightforward generalization to any 3d $O(N)$ model by simply replacing the critical indices with the appropriate values. Table \ref{tab:1} shows our results for $N=0$ (the non-unitary self-avoiding walk model), $N=1$ (Ising), $N=2$ ($XY$ model) and $N=3$ 
(Heisenberg model), where we can compare our results with the two-loop 
calculation of \cite{Diehl:1998mh}.

\subsection{The extraordinary transition.}
Such a transition is characterized by the non-vanishing contribution of the boundary identity to the two-point functions of $\mb{Z}_2$ odd operators. 
In this case the boundary surface is in an ordered phase, 
therefore the degrees of freedom described by $\mb{Z}_2$ odd operators are frozen. The first non-vanishing surface operator, besides the identity, is the displacement $D$ with $\Delta_{D}=3$. As a consequence, the most relevant contribution to the boundary channel is known and the crossing equations can be exploited to obtain information on the bulk channel.

Actually adding the boundary identity to the truncation requires adding more bulk operators as well. We found a first stable solution of the type (4,1,1).  This time the 
scaling dimensions of the two needed bulk scalars 
$\varepsilon''$ and $\varepsilon'''$ cannot be used as input parameters because, once fixed $\Delta_\sigma$, $\Delta_\varepsilon$ and 
$\Delta_{\varepsilon'}$\footnote{Here and in the rest of this section we use as input parameters of the Ising model the values $\Delta_\sigma=0.518154(15)$, $\Delta_\varepsilon=1.41267(13)$ and $\Delta_{\varepsilon'}=3.8303(18)$ taken from \cite{El-Showk:2014dwa}.}, we get a solution only if
\beq
N=1\,,\qquad\Delta_{\varepsilon''}=7.316(14)\,,\qquad
\Delta_{\varepsilon'''}=13.05(4)\footnote{In the entire paper the
  estimate of the statistical error due to the uncertainty on the
  input parameters is obtained by means of a statistical bootstrapping procedure.}.
\label{e411}
\eeq 
The other parameters of the solution are
\begin{align}
a_\varepsilon\lambda_{\sigma\sigma\varepsilon} =
6.914(6), &&a_{\varepsilon'}\lambda_{\sigma\sigma\varepsilon'}=2.261(2), &&
a_{\varepsilon''}\lambda_{\sigma\sigma\varepsilon''}=0.187(1),
\notag \\
a_{\varepsilon'''}\lambda_{\sigma\sigma\varepsilon'''}=0.0046(1), &&a_\sigma^2=6.757(4)\,,
&& \mu^2_{\si D}/C_D= 0.06282(3); 
\label{ecoupling}
\end{align}
where we denoted with $C_D$ the Zamolodchikov norm of the displacement operator.

%We can easily generalize this solution to the $O(N)$ spin models using as input 
%the data of table \ref{tab:1}. For the $XY$ model $(N=2)$ we obtain
%\beq
%N=2\,,\qquad\Delta_{\varepsilon''}=6.002(33)\,,\qquad
%\Delta_{\varepsilon'''}=10.96(4).
%\label{xy411}
%\eeq
%with
%\begin{align}
%a_\varepsilon\lambda_{\sigma\sigma\varepsilon} =
%5.585(11), &&a_{\varepsilon'}\lambda_{\sigma\sigma\varepsilon'}=1.466(10), &&
%a_{\varepsilon''}\lambda_{\sigma\sigma\varepsilon''}=0.307(9),
%\notag \\
%a_{\varepsilon'''}\lambda_{\sigma\sigma\varepsilon'''}=0.0162(3), &&a_\sigma^2=5.495(9)\,,
%&& \mu^2_{\si D}/C_D= 0.06741(7).
%\label{xycoupling}
%\end{align}

%Similarly for the Heisenberg model $(N=3)$ we get
%\beq
%N=3\,,\qquad\Delta_{\varepsilon''}=5.285(17)\,,\qquad
%\Delta_{\varepsilon'''}=10.48(1).
%\label{h411}
%\eeq
%with
%\begin{align}
%a_\varepsilon\lambda_{\sigma\sigma\varepsilon} =
%5.019(8), &&a_{\varepsilon'}\lambda_{\sigma\sigma\varepsilon'}=0.846(13), &&
%a_{\varepsilon''}\lambda_{\sigma\sigma\varepsilon''}=0.505(10),
%\notag \\
%a_{\varepsilon'''}\lambda_{\sigma\sigma\varepsilon'''}=0.0207(1), &&a_\sigma^2=4.874(8)\,,
%&& \mu^2_{\si D}/C_D= 0.06919(10).
%\label{hcoupling}
%\end{align}

%It is interesting to notice that  the value
%of $\Delta_{\varepsilon''}=3+\omega_2$ as a function of $N$ is close to 
%that first calculated 
%in \cite{Litim:2002cf} with the functional renormalization group method.

\begin{figure}
\centering
\includegraphics[width=11 cm]{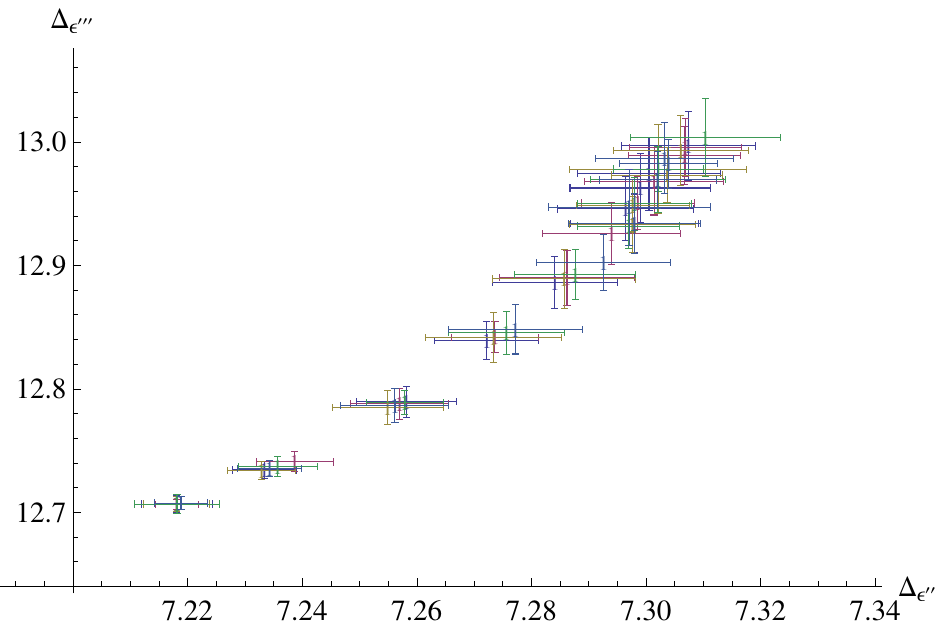}
\caption{Parametric plot of the  scaling dimensions of  
$\Delta_{\varepsilon''}$ and $\Delta_{\varepsilon'''}$ generated by the unknown parameter $\widehat\Delta$ in the (4,2,1) truncation. Here we see the 
effect of the statistical errors on the input data, namely $\Delta_\sigma$,
$\Delta_{\varepsilon}$ and $\Delta_{\varepsilon'}$ as well as the 
effect of the spread of the solutions. 
Some of these data are presented in table \ref{tab:3}.}
\label{fig:2}
\end{figure}

%In the case of the Ising model, where we used the  more precise input data of 
% \cite{El-Showk:2014dwa}, 
In this case, we probed the stability of the solution by adding a new conformal block in the boundary channel. It turns out that the truncation (4,2,1) defines a one-dimensional family of the solutions, where the free parameter is the dimension of the added surface operator, which can vary in the range 
$0<\widehat\Delta\le\infty$. In the limit $\widehat\Delta\to\infty$ we recover, as expected in a stable solution, the truncation (4,1,1). The dimensions of the two bulk operators $\Delta_{\varepsilon''}$ and $\Delta_{\varepsilon'''}$ vary as functions of $\widehat\Delta$ in a narrow range: the net effect of the unknown parameter is to reduce a bit the scaling dimensions of these bulk operators. Eliminating $\widehat\D$ we obtain the plot in fig. 
\ref{fig:2}.      
The uncertainty on the actual value of  $\widehat{\Delta}$ forces us to enlarge the errors in the bulk dimensions. Fig. \ref{fig:2} roughly suggests 
\beq
\Delta_{\varepsilon''}=7.27[5]\,,\qquad
\Delta_{\varepsilon'''}=12.90[15],
\label{e421}
\eeq 
which supersede eq. \eqref{e411}. We used  square brackets to indicate that this is not a statistical error, but a sum of the uncertainties.

Unfortunately one can find in literature   a wide range of proposed 
values for $\Delta_{\varepsilon''}$ and $\Delta_{\varepsilon'''}$ which strongly 
depend on the method employed (see for instance table 3 of \cite{El-Showk:2014dwa}). What is especially disturbing for us is that the method of determinants applied to the four-point function gave very different values for these quantities \cite{Gliozzi:2014jsa}, so we decided to reanalyze the bootstrap 
equations for the four-point function on the bulk in order to see whether there is also a solution compatible with the spectrum suggested by the boundary bootstrap. Out of this study we can confirm the existence 
of a scalar of dimension $\sim7.2$ with a positive coupling. We were unable to 
find a proper solution for the scalar at $\sim13$, all solutions being 
characterized by a coupling that is very small, negative and nearly always compatible with zero. The quoted  dimensions of these two scalars found with the linear functional method \cite{El-Showk:2014dwa} are respectively $\sim7$ and $\sim10.5$.

Another interesting two-point function to be studied in the 
extraordinary transition of the Ising model  is the spin-energy correlator $\braket{\sigma(x)\varepsilon(y)}$ which is different form zero only in this phase,
being the only surface transition where
 the $\mb{Z}_2$ symmetry of the model is broken. 
The fusion rule of the bulk sector contains odd operators only:
\beq
\sigma\times\varepsilon\sim\, \sigma+\sigma'+\sigma''+\dots,
\label{sigmaepsilonBulk}
\eeq
while in the boundary sector the first primary operator contributing, besides the identity, is the displacement operator:
\beq
\sigma \sim\,1+D+\dots,\qquad \varepsilon \sim\,1+D+\dots
\label{sigmaepsilonBdy}
\eeq
The first stable solution corresponds to the truncation
(3,1,1) defined by the above fusion rules. It is associated with the (apparently) common intersection of the zeros of the $5\times5$ determinants made with 
the derivatives of the 5 conformal blocks involved (see fig. \ref{fig:3}):
\begin{figure}
\centering
\includegraphics[width=9 cm]{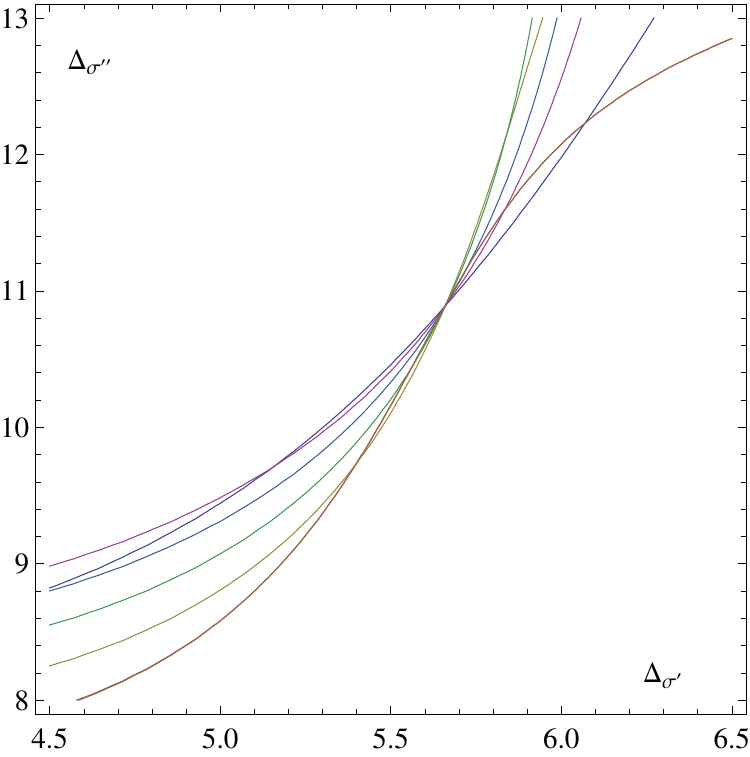}
\caption{Plot of the zeros of some $5\times5$ determinants associated with the fusion rules (\ref{sigmaepsilonBulk}) and  (\ref{sigmaepsilonBdy}).}
\label{fig:3}
\end{figure}

\beq
\Delta_{\sigma'}\simeq5.66~; \qquad \Delta_{\sigma''}\simeq10.89~;
\eeq

\beq
a_\sigma 
\lambda_{\sigma\varepsilon\sigma} \simeq0.148\,\kappa~; \quad 
a_\varepsilon a_\sigma\simeq0.927\,\kappa~; \quad
\m_{\sigma D} \m_{\varepsilon D}/C_D\simeq0.0196\,\kappa~.
\label{emixcoupling}
\eeq  
The parameter $\kappa$ arises because now the bootstrap equations are homogeneous, that is, they do not contain the information about the normalization of the external operators. The normalization of the order parameter is contained in the correlator $\braket{\si \si}$, while the normalization of the energy follows from assuming symmetry of the OPE coefficient $\lambda_{\sigma\sigma\varepsilon}~=~\lambda_{\sigma\varepsilon\sigma}$. Therefore, combining (\ref{emixcoupling}) with the analogous couplings in (\ref{ecoupling}), we can compute
the unknowns 
$a_\varepsilon,\ a_\sigma,\ \m_{\sigma D}/\sqrt{C_D},\ \m_{\varepsilon D}/\sqrt{C_D},\ \kappa,\ 
\lambda_{\sigma\varepsilon\sigma}$. 

In order to probe the stability of the solution and to evaluate the errors we upgraded the solution to (5,1,1), which corresponds to a one-parameter family of solutions. We used as a free parameter the heaviest bulk scalar $\si_4$. A solution exists for $18\le\D_{\si_4}\le28$. As expected for a stable solution, this parameter has no visible effect on the OPE coefficients and only slightly affects the scale dimensions of the two scalar $\si''$ and $\si'''$.  The results of this analysis  can be found in table \ref{tab:2} 

\begin{table}[htb]
\centering
\begin{tabular}{*{6}{>{$}c<{$}}}
\toprule
\D_{\varepsilon''} &\D_{\varepsilon'''} &\D_{\si'} &\D_{\si''}&\D_{\si'''}&\lambda_{\si\si\varepsilon} \\
\midrule
7.27[5]   &  12.90[15]   & 5.49(1) &  10.6[3]&16[1]&1.046(1) \\
\bottomrule
\vspace{0.005\textheight}
\end{tabular}
\centering
\begin{tabular}{*{4}{>{$}c<{$}}}
\toprule
a_{\varepsilon}&\mu_{\varepsilon D}/\sqrt{C_D}&a_{\si}&\mu_{\si D}/\sqrt{C_D}\\
\midrule
6.607(7)&1.742(6)&2.599(1)&0.25064(6)\\
\bottomrule
\end{tabular}
\caption{The main results of the combined analysis of $\braket{\si\si}$ and
$\braket{\si\varepsilon}$ in the extraordinary transition are split in two parts. The top table refers to data of the bulk channel, while the bottom table contains OPE coefficients specific to the boundary channel of the extraordinary transition. Errors in square brackets refer to data whose uncertainties depend 
on an  unknown parameter; the other errors simply reflect the statistical errors of the input data, namely, $\D_\si$, $\D_{\varepsilon}$ and $\D_{\varepsilon'}$. }
\label{tab:2}
\end{table}

It turns out that $\Delta_{\sigma'}$ is nicely close to the bound $\Delta_{\sigma'}\le 5.41(1)$ found in \cite{Kos:2014bka}.
 Notice also that the resulting OPE coefficient $\lambda_{\sigma\sigma\varepsilon}$ 
is in perfect agreement with the estimate of a recent Monte Carlo calculation
\cite{Caselle:2015csa} which gives
$\lambda_{\si\si\varepsilon}=1.07(3)$ and the value  
$(\lambda_{\si\si\varepsilon})^2=1.10636(9)$       found in 
\cite{El-Showk:2014dwa} through the study of the four-point function with the linear functional method.

There is another very impressive check of these results.
The Ward identity associated with the displacement operator tells us that 
the quantity $x_O=\Delta_O\frac{a_O}{\mu_{OD}}\sqrt{C_D}$  does not depend on the 
specific bulk operator $O$ but only on the surface transition, as described in section \ref{sec:interface}. The above results yield 
\beq
x_\sigma=5.3727(27)~; \quad x_\varepsilon=5.358(15)~,
\eeq
showing, within the errors, a reassuring fulfillment of the  Ward identities.

\paragraph{Note added, November 2021} A previous version of this paper contained results about the extraordinary transition for $N>1$. However, the first operator in the boundary channel was incorrectly assumed to be the displacement. Instead, a protected boundary operator of dimension $\wh{\D}=2$ arises from the breaking of the continuous $O(N)$ symmetry. We refer to \cite{Padayasi:2021sik} for a conformal bootstrap study of this boundary condition for $N>1$.

\begin{figure}
\centering
\includegraphics[width=9 cm]{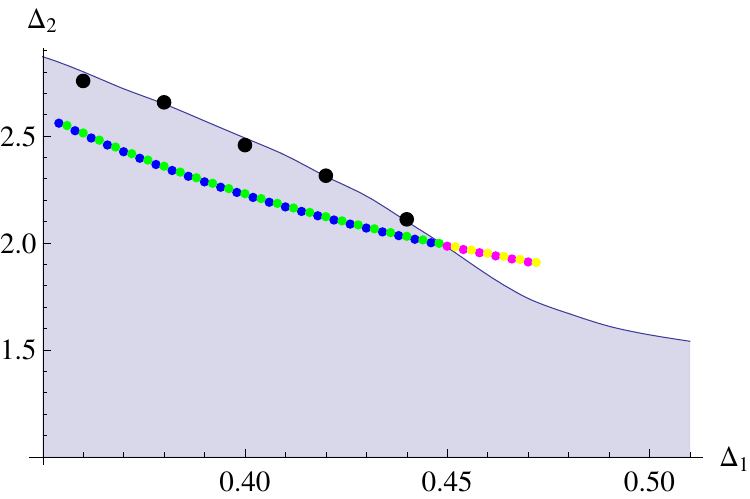}
\caption{Plot of the one-parameter family of the truncation (4,3,0) 
in the plane $(\widehat{\D}_1,\widehat{\D}_2)$, superimposed to the 
upper unitarity bound found in\cite{Liendo:2012hy}. 
The blue and green dots correspond respectively 
to the minimal and the maximal choice  of the pair $(\Delta_{\varepsilon''},
\Delta_{\varepsilon'''})$, as determined in fig. \ref{fig:2}. These dots are replaced by ones respectively magenta and yellow when some OPE coefficient become negative. For the black dots on the unitarity bound see explanation in the text.}
\label{fig:4}
\end{figure}

\subsection{The special transition.}
According to our discussion at the beginning of this section, solutions 
ascribed to the special transition are associated with truncations of the form
 $(m,n,0)$ in which all the OPE coefficients are non-negative. By consistency with the results of the previous subsection we have to use the same bulk spectrum 
determined in the extraordinary transition. We found solutions of the form (3,3,0) and (4,3,0) with similar properties. Here we only discuss  the latter.

Instead of an isolated solution, in this case we find a one-parameter family in the three-dimensional space of the boundary scale dimensions 
$(\widehat{\D}_1<\widehat{\D}_2<\widehat{\D}_3)$. The lowest-dimensional operator has to be identified 
 with $\widehat\sigma$ and according with the two-loop calculation of  \cite{Diehl:1998mh} we 
expect $\widehat{\D}_{\widehat\sigma}\sim~0.42$. In our case a unitary 
solution exists only for $0.34\le\widehat{\D}_1\le0.45$. Below 0.34 the solution disappears abruptly; above 0.45 it becomes non-unitary.

\begin{figure}[h!]
\centering
\includegraphics[height=0.2\textheight,width=0.47\columnwidth]{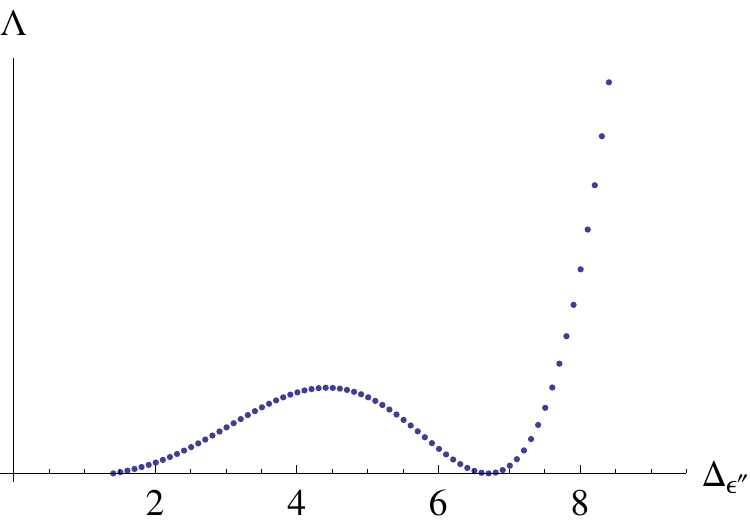}
\includegraphics[height=0.2\textheight,width=0.47\columnwidth]{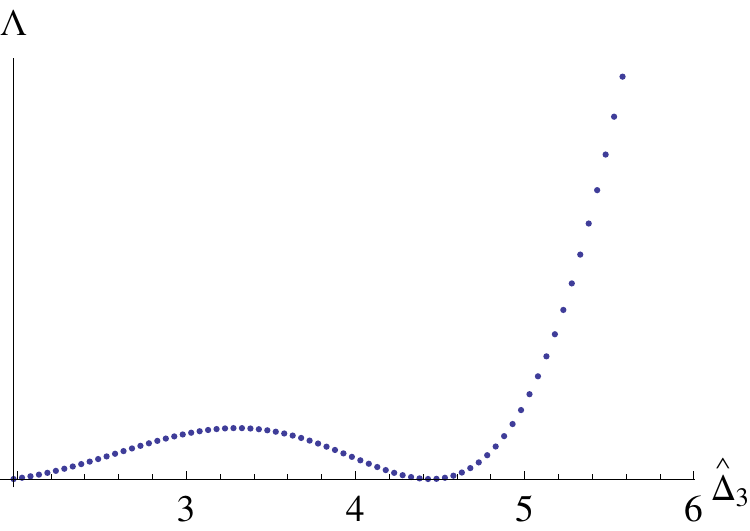}
\caption{Linear functionals for the bulk and boundary channels in the special transition.}
\label{fig:functionals}
\end{figure}

Using $\widehat{\D}_3$ as a free parameter, we obtain the plot of fig. \ref{fig:4}, which is superimposed to the unitarity upper bound found in 
\cite{Liendo:2012hy}. As expected,  the transition 
to the non-unitary region coincides with the unitarity boundary found 
by the linear  functional method.
Consistency requires  that the spectrum of our solution 
at the intersection should agree with the one extracted from the zeros 
of the  linear functional\cite{ElShowk:2012hu} calculated
 at the same point. In fact, the first zero of the linear functional at the 
intersection point, in the bulk sector, is (see fig. \ref{fig:functionals})
around $\sim 6.7$, which is consistent with our result 
for $\Delta_{\varepsilon''}$. Similarly, the zero of 
the extremal functional for the boundary sector (besides $\widehat{\D}_1$ and 
$\widehat{\D}_2$) is perfectly consistent with the 
 value  $\widehat{\D}_3\sim 4.44$ at the crossing point.
 
Such a  boundary required by unitarity could also be seen as the locus were one or more OPE coefficients change sign.  Our solution leads us to conjecture that the couplings 
vanishing at the unitarity bound are $\lambda_{\sigma\sigma\varepsilon'}$ and  
$\lambda_{\sigma\sigma\varepsilon'''}$. In the construction of the 
upper unitarity bound in \cite{Liendo:2012hy} it is assumed that the 
first bulk primary is the Ising energy $\varepsilon$ and it follows that the subsequent primary has scale dimension larger than $\Delta_{\varepsilon'}$, as suggested by our conjecture.  

The knowledge of the linear functional leading to the bound of fig. \ref{fig:4} suggests another interesting  cross-check  of the two methods: 
given a value of $\widehat{\D}_1$ we  insert in the (4,3,0) truncation the 
first four zeros of the linear functional on the bulk channel and evaluate with the method of determinants the corresponding boundary values $\widehat{\D}_2$ and  $\widehat{\D}_3$. 
It turns out that in the plane $\widehat{\D}_1,\widehat{\D}_2$ such a solution   lies on the unitarity bound, as consistency requires (see black dots in fig. \ref{fig:4}).

\section{Renormalization group domain wall for the $O(N)$ model.}
\label{sec:interface}

Before starting the exploration of a specific conformal interface, let us recall the relevant CFT data that one needs to collect in order to completely describe the generic system. Conformal interfaces are closely related to boundaries. In fact,
as we mentioned in section \ref{sec:method}, an interface between a CFT$_1$ and a CFT$_2$ can be mapped to a boundary problem using the folding trick. One turns the original setup into a boundary for the theory CFT$_1\times \overline{\textup{CFT}}_2$, where the bar means that a reflection $x^d\to-x^d$ has been applied to one of the theories. We see that the natural bulk CFT data is given by the value of the two point functions of operators placed in mirroring points with respect to the interface: they are mapped to expectation values of operators in the folded CFT. This also identifies the needed operators as primaries of the folded theory, which in particular include all bulk primaries of the two CFTs. The latter are not sufficient, though, because they do not play any role as building blocks of correlators across the interface. Another way of understanding this circumstance is provided by the north-south pole quantization, or equivalently by conformally mapping the theory to a $d$-dimensional sphere. Local operators at the north or south pole create a state belonging to the Hilbert space of either CFT. The interface is a linear map between the Hilbert spaces, and the correlators of operators placed in mirroring points - that is, at the north and south poles - are the matrix elements of this map. Analogous considerations are valid for the bulk-to-defect couplings. Let us now turn to the specific interface we shall study in this paper.

The \emph{Renormalization group domain walls} are interfaces between two CFTs which lie at the top and at the bottom of an RG flow. More precisely, there is an easy operational definition: start with a CFT on the whole space, and modify the action by integrating a relevant operator over half of the space. Far away in this region, the long distance physics will be dominated by the CFT  at the bottom of the flow triggered by the perturbation. This definition can be employed literally when the coupling is only mildly relevant, and perturbation theory makes sense. In order to single out a unique gluing condition, it is also necessary to specify which defect deformations are turned on along with the bulk flow. In the case of interest for us, we shall argue that no marginal deformations exist on the defect, and so we just choose to fine tune perturbatively the relevant defect couplings.  As usual, near the interface the critical behaviour is modified with respect to both the UV and the IR homogeneous fixed points, with new critical exponents arising. RG domain walls have been mainly studied in two dimensions \cite{Fredenhagen:2005an,Brunner:2007ur,Gaiotto:2012,Konechny:2014opa,Poghosyan:2014jia}. In a general non perturbative setting, the determination of the defect spectrum and the computation of correlators is a very difficult task. In some limiting cases, however, some of the answers might be found with little effort. For instance, a relevant operator may force the bulk to flow towards a trivial theory. In this case, the RG interface is reduced to a boundary condition for the ultraviolet CFT. As an example, consider giving a mass to a free boson on half of the space, in any dimension greater than two. Correlators on the perturbed side are exponentially damped, and at large distances the theory is empty. From an RG point of view, the coupling grows in the IR, and the configurations of non-zero field on the perturbed side are suppressed in the partition function. As a consequence, a Dirichlet boundary condition is imposed to the massless free boson on the other side. 

A more interesting case is the RG domain wall corresponding to the Wilson-Fisher fixed point of the $O(N)$ model with $(\p^2)^2$ interaction. This interface is captured by the following bare action:
\beq
\mc{S}= \int\!\!d^dx\, \frac{1}{2S_d (d-2)}\,\pa_\m \p^i\,\pa_\m \p^i+\th(x^d) \frac{g}{4!}{(\p^i\p^i)}^2,
\label{eq:RGaction}
\eeq
where $\th(x^d)$ is the Heaviside function, $S_d=2\pi^{d/2}/\,\Gamma\!\left(d/2\right)$ and we chose to normalize the elementary field so that it has a canonical two-point function in free theory. 
As we pointed out, a question that needs to be answered concerns the stability of this interface. One needs to know how many relevant operators must be fine-tuned, and if marginal deformations exist. The interface possesses a weakly coupled description in $4-\ep$ dimensions, and, at the classical level, the only relevant defect primary in the singlet sector is $\widehat\p^2$. Once we tune it to zero, unlike the situation in the special transition, we do not impose Neumann boundary conditions, but only continuity of $\pa_z\widehat \p^i$ on the interface. Hence, the classically marginal operator $\pa_z\widehat\p^2$ does not vanish, and should be taken into account. We shall show that this operator becomes irrelevant at one loop. Therefore, the RG interface appears to be isolated in perturbation theory.

In the following, we characterize the correlations of scalar primaries in the presence of the domain wall at lowest order in $\ep$-expansion. Along the way, we point out that correlations across the interface encode at this order the mixing induced by the RG flow among nearly degenerate operators \cite{Gaiotto:2012}. This is true in the larger class of perturbation interfaces constructed by means of a nearly marginal deformation. We then focus on the RG domain wall between the three dimensional free theory and the Ising model, and study the two-point function of the field $\si$ using the method of determinants. We also provide some non-perturbative information on generic conformal interfaces involving the free theory, by noticing that some of the crossing constraints can be solved analytically.

\subsection{The $\ep$-expansion and the role of the displacement operator.} 

Since the UV side of this RG interface is a free theory, the interface itself is not captured by mean-field theory: the CFT data related to it is $O(\ep)$ in perturbation theory. One can easily obtain general results at leading order by exploiting the Ward identity eq. \eqref{eq:stressWard}, which defines the displacement operator. The identity tells us that we can move the interface in the orthogonal direction by integrating the displacement in the action. Its insertion in a correlation function is therefore equivalent to a derivative with respect to the position of the interface, that is,
\beq
\int\!\! d^{d-1}y\, \braket{D(y^a)\, O_1(x_1)\dots O_n(x_n)} = 
-\sum_{i=1}^n \frac{\pa}{\pa x_i^d} \braket{O_1(x_1)\dots O_n(x_n)}.
\label{eq:integratedWard}
\eeq
Since the violation of translational invariance happens at order $g$ - see eq. \eqref{eq:RGdisp} - the relation \eqref{eq:integratedWard} rephrases some information about an $n$-point function of order $g^L$ in terms of the integral of a $(n~+~1)$-~point function of order $g^{L-1}$. In general, knowledge of the variation with respect to the position of the interface is obviously insufficient for reconstructing the full correlator. However, all configurations  of two points are conformally equivalent to the one in which the points are aligned on a line perpendicular to the defect. Therefore a two-point function can be traded for the integrated three-point function on the l.h.s. of eq. \eqref{eq:integratedWard}. The advantage is that the integral does not generate additional divergences: one only needs to renormalize the theory at order $g^{L-1}$. On the other hand, it is still necessary to determine a primitive of the l.h.s. of eq. \eqref{eq:integratedWard} as a function of the position of the interface. We shall see that this is possible at lowest order: the tree level $2$-point correlator, which is just the homogeneous one, can be used to compute the one loop correction in the presence of the interface. 

It is simple to derive from \eqref{eq:integratedWard} a new scaling relation. As pointed out, when two operators are placed in mirroring points, in which case $\xi = -1$, their correlator is equivalent, through the folding trick, to a one-point function:
\beq
\braket{O_\textup{L}(x)\,O_\textup{R}(\mc{R}x)}=
\frac{a_{\textup{L\,R}}}{\abs{2x^d}^{\D_\textup{L}+
\D_\textup{R}}}, \qquad \mc{R}x=\left(x^a,-x^d\right).
\label{eq:twopmirror}
\eeq
Here we think of $O_\textup{L}$ and $O_\textup{R}$ as scalars belonging respectively to the UV and IR spectrum. Similarly, the three-point function $\braket{O_\textup{L}\,O_\textup{R}\, D}$ is fixed up to a number:
\beq
\braket{O_\textup{L}(x)\,O_\textup{R}(\mc{R}x)\, D(y^a)}=
\frac{\mu_{\textup{L\,R}\,D}}{\abs{2x^d}^{\D_L+\D_R-d}\,\abs{x-y}^{2d}}
\eeq
Using the fact that in this geometry $\xi$ is stationary with respect to orthogonal displacements of the interface, it is easy to derive the following relation between these pieces of CFT data
\beq
\frac{(\D_\textup{R}-\D_\textup{L})a_{\textup{L\,R}}}{S_d}
=\mu_{\textup{L\,R}\,D}.
\label{eq:scalingAcross}
\eeq
In the particular case where one of the bulk operators is the identity, one recovers a relation which was first noticed in the case of a boundary by Cardy \cite{Eisenriegler} (see also \cite{McAvity:1995zd}):
\beq
\pm\,\frac{\D_k a_k}{S_d}=\mu_{k D},
\label{eq:scalingOnepoint}
\eeq
where the plus/minus sign is valid for the interacting/free side respectively. We start by using eq. \eqref{eq:scalingOnepoint} to determine the $a_k$'s. The answer at order $\ep$ is quite simple: only one operator acquires expectation value, on both sides of the interface. To see this, let us identify the displacement. Looking at the action \eqref{eq:RGaction}, we see that the interface is displaced at leading order by integrating the bare operator $g (\p^2)^2/4!$, that is\footnote{Notice that at higher orders the interacting stress-tensor needs to be improved to be kept finite and traceless \cite{Brown:1979pq}. The improvement is proportional to $(\pa_\m\pa_\n-\de_{\m\n}\pa^2)\p^2$, so that the displacement receives a contribution from the operator $\pa_a\pa^a\p^2$.}
\beq
D= \frac{g}{4!} (\p^2)^2 +\mc{O}(g^2)
=\frac{1}{8(N+8)\pi^2}\ep\,(\p^2)^2+\mc{O}(\ep^2),
\label{eq:RGdisp}
\eeq
where we plugged the fixed point value of the coupling at order $\ep$: 
\beq
g^*=\frac{3}{(N+8) \pi^2}\,\ep.
\label{eq:gstar}
\eeq
Now, since $(\p^2)^2$ is a primary of the free theory, and no other primary mixes with it at order one, its correlation function with any other primary is zero at leading order. This means that all coefficients $\m_{OD}=\mc{O}(\ep^2)$, but for the case $O=(\p^2)^2$. Using the relation \eqref{eq:scalingOnepoint}, we conclude that the only non vanishing expectation value at this order is $\braket{(\p^2)^2}$. We can then obtain the number $a_{\p^4}$ at order $\ep$ from a tree level computation. Indeed, the relevant bulk-to-defect coupling is given at leading order by
\beq
\m_{\p^4 D}=|x|^{8}\braket{\,\frac{(\p^2)^2(x)}{\sqrt{8N(N+2)}}\,D(0)} \\
= \frac{\sqrt{2N(N+2)}}{4(N+8)\pi^2}\,\ep.
\eeq
Therefore
\beq
a^\textup{IR}_{\p^4} = - a^\textup{UV}_{\p^4} = \frac{\sqrt{2N(N+2)}}{8(N+8)}\,\ep.
\label{eq:aUVaIR}
\eeq
Let us make a comment. It was obvious from the start that only a small class of operators could exhibit a one-point function at first order in the coupling: four powers of the elementary field are needed to contract a single vertex, and of course the operator must be in the singlet of $O(N)$. However, infinitely many scalar primaries can be constructed in free theory which fulfill these requirements, involving an increasing number of derivatives of the fields\footnote{That these primaries must exist can be seen independently from their expression in terms of elementary fields, for instance from the asymptotics of the two point function of $\p^2$ in a free theory with a boundary. The presence of the identity in the boundary channel can only be balanced by an infinite number of conformal blocks in the bulk channel. Only one primary can be built with two powers of the fields, so the rest are the ones we are interested in. The explicit conformal block decomposition for this case can be found in \cite{Liendo:2012hy}. It is also amusing to notice that, analogously to the case at hand, this tower of operators does not contribute at order $\ep$ to the two-point function of $\p$ with Dirichlet or Neumann boundary conditions. As noticed in \cite{Liendo:2012hy}, in that case the OPE coefficients $\la_{\p\p\,\pa^{2k}\!\p^4}$ are the vanishing quantities at order $\ep$. 
}. 
The simplest use of eq. \eqref{eq:aUVaIR} is the determination of the most general two-point function of operators lying on the same side of the interface at order $\ep$. Sticking for simplicity to the case of external scalars, one simply writes
\begin{multline}
\braket{O_1(x)O_2(x')} \\
=\frac{1}{(2x_d)^{\D_1}(2x'_d)^{\D_2}}\,\xi^{-(\D_1+\D_2)/2}
\left(\de_{12}+\la_{12\p^4}a_{\p^4} \fk^{d=4}(\D_{12},\D=4,\xi)\right)
+\mc{O}(\ep^2).
\label{eq:2pointsame}
\end{multline}
Notice that $\la_{12\p^4}$ is guaranteed to belong to the 4d free theory only when $O_1$ and $O_2$ are on the UV side. Indeed, primaries on the interacting side are in general a mixture of classically degenerate renormalized operator, and when the mixing happens at leading order $\la_{12\p^4}$ becomes a linear combination of UV OPE coefficients. For completeness, we compare this derivation with some direct one loop computations in appendix \ref{sec:details}.  

As pointed out in the introduction to this section, in order to capture correlations across the interface we would need all the one-point functions of the folded theory. This set encompasses the $a_{L\,R}$ defined in \eqref{eq:twopmirror}, and is much bigger. It is in fact more viable to reach for the two-point functions of primaries directly through the integrated Ward identity eq. \eqref{eq:integratedWard}, specified to the case of interest:
\beq
\int\!\! d^{d-1}y\, \braket{O_\textup{L}(x)\,O_\textup{R}(x') D(y)} = 
-\left(\frac{\pa}{\pa x^d}+\frac{\pa}{\pa x'^d}\right) \braket{O_\textup{L}(x)\,O_\textup{R}(x')}.
\label{eq:2pointWard}
\eeq
We pick for the left hand side the three-point function of primaries in the translational invariant theory, and we get the one-loop two-point function by integrating over the position of the displacement. Notice that in doing so we disregard the mixing of primaries with descendants. In the cases in which this happens at order one, on the left hand side of eq. \eqref{eq:2pointWard} additional terms needs to be taken into account, which have the form of a three-point function involving derivatives of a primary operator. Consider first two operators which are degenerate in the free theory. In other words,
\beq
\D_{\textup{LR}}\equiv \D_\textup{L}-\D_\textup{R}=\mc{O}(\ep).
\eeq
In this case eq. \eqref{eq:2pointWard} can only be used to determine the one loop correlator up to a constant. Indeed, since both $\m_{\textup{L\,R}\,D}$ and $\D_\textup{L}-\D_\textup{R}$ are of order $\ep$, one needs the one loop three-point function to determine $a_{\textup{L\,R}}$ from eq. \eqref{eq:scalingAcross}. This is the familiar effect of degeneracies in perturbative computations, and is related to the mixing of operators along the RG flow (see section \ref{subsec:mixing}). Integration of \eqref{eq:2pointWard} is straightforward, and one gets
\beq
\braket{O_\textup{L}(x)\,O_\textup{R}(x')}= \frac{a_\textup{L\,R}}{\abs{2x^d}^{\D_\textup{L}}(2x'^d)^{\D_\textup{R}}}(-\xi)^{-\D_\textup{L}}
\left(1+\frac{\D_\textup{LR}}{2} \log (-\xi) \right),\quad
\D_\textup{LR}=\mc{O}(\ep).
\label{eq:acrossdeg}
\eeq
Comparing with the form \eqref{eq:twoptscalar} we can write at this order
\beq
G_\textup{L\,R}(\xi) = a_\textup{L\,R}
\qquad\D_\textup{LR}=\mc{O}(\ep).
\eeq
A comment is in order. The presence of a logarithmic singularity compatible with exponentiation is somewhat natural, since turning the coupling off one recovers the short distance power low divergence proper of the homogeneous theory. However, there is no reason for this to happen when considering the OPE limits in the Euclidean defect CFT. The exponentiation agrees in the large $\xi$ limit with the defect OPE, as it is easy to verify using the formulae given in subsection \ref{subsec:mixing}. On the other hand, no small $\xi$ limit exists for primaries on opposite sides of the domain wall, and in fact the folded cross-ratio is $\xi_\textup{folded}=-(1+\xi)$, which vanishes when the operators are placed in mirroring points. We decide to keep using the form \eqref{eq:twoptscalar}, and notice that it might be fruitful to look for a justification in Lorentzian signature, where the small $\xi$ limit corresponds to light-like separated operators.

In the case of operators with dimension differing in the UV limit, the two-point functions at one loop can be fixed completely. Due to $O(N)$ and rotational symmetry, $\D_{\textup{LR}}$ is an even integer in $d=4$, which provides a simplification. The computation is slightly more involved than in a previous case, and we give some details in appendix \ref{sec:details}. The result in the case $\abs{\D_{\textup{LR}}}=2$ is different from all the others:
\beq
G_\textup{L\,R}(\xi) = \frac{\pi^2}{2} \m_{\textup{L\,R}\,D}\, \textup{sign} (\D_\textup{LR}) \,   (\xi-1),
\qquad \abs{\D_\textup{LR}}=2,
\label{eq:across2}
\eeq
while
\begin{multline}
G_\textup{L\,R}(\xi)  = -\frac{\pi^2 \Gamma (2 k+3)}{(k-1) k^2 \Gamma (k+2)^2} \m_{\textup{L\,R}\,D}\, \textup{sign} (\D_\textup{LR}) \\
 (-\xi)^{k+1} \left\{\Big( (4 k+2) \xi ^2+3 (k+1) \xi +1\Big)\, _2F_1\left(-k-1,-k,-2 (k+1);\frac{1}{\xi
   }\right) \right. \\
\left. -  \Big((4 k+2) \xi^2 +(k +2)\xi \Big) \, _2F_1\left(-k-1,-k-1,-2(k+1);\frac{1}{\xi }\right)\right\}, \\
\qquad \abs{\D_\textup{LR}}\equiv2k>2.
\label{eq:across>2}
\end{multline}

As one might have expected, the hypergeometric functions in eq. \eqref{eq:across>2} are in fact polynomials.
 
These results complete the analysis of bulk correlations at order $\ep$, if knowledge of the $\la_{123}$ is assumed: $n$-point functions of bulk operators are determined by taking successive OPEs on the two sides until one is left with a one-point function or a two-point function across the interface. We shall content ourselves of this leading order solution, but we would like to comment on the possibility of generalizing the procedure. Unfortunately, the number of non vanishing one-point functions is infinite already at next to leading order\footnote{This statement again follows immediately from the fact that the operator $\p^2$ acquires an expectation value at order $\ep^2$.}. Therefore, once the displacement has been correctly normalized, one has to compute the relevant three-point functions at one loop and integrate them to find the two loop two-point functions.

Let us now consider the defect spectrum at order $\ep$. The dimensions of the operators can be extracted through the defect OPE decomposition of eq. \eqref{eq:2pointsame}. When nearly degenerate operators are present in the UV theory, also the defect operators mix, and the spectrum is given by the eigenvalues of the matrix of anomalous dimensions. We shall deal with this more general case in the next subsection. Here we comment on some features of the spectrum focusing for simplicity on the non-mixing operators. The lightest defect scalar in the OPE of a bulk operator $O$ has dimension
\begin{align}
\wh{\D}_O &= \D^\textup{UV}_{O}-2\la_{OO\p^4}\,a^\textup{UV}_{\p^4}+\mc{O}(\ep^2) \notag \\
        &= \frac{1}{2} (\D^\textup{UV}_{O}+\D^\textup{IR}_{O})+\mc{O}(\ep^2).
\label{eq:interfSpectrum}
\end{align}
The second equality in eq. \eqref{eq:interfSpectrum}, which agrees with first order conformal perturbation theory, says that the defect primary stands half way between the corresponding infrared and ultraviolet operators in the bulk. Let us make some more specific comments. $\wh{\D}_{\p^4}=4-\ep$ is the protected dimension of the displacement operator. This is expected, even if there are degenerate operators in free theory. Two primaries exist with dimension near to four, but both of them are protected, the second one being the displacement of the folded theory. The second interesting scale dimension is obtained by going one step further in the defect OPE of $\p^2$. We encounter the operator $\pa_z \widehat\p^2$, and since no other scalars exist which could mix with it, we can safely read off his dimension from the boundary block decomposition: $\wh{\D}_{\pa\p^2}=3-\frac{N+14}{2(N+8)}\ep$. We see that this scalar is irrelevant at the Wilson-Fisher fixed point, so that the stability of the interface is not altered by its presence. A third remark concerns the odd spectrum. Since the anomalous dimension of $\p^i$ starts at two loops, or equivalently the bulk OPE does not contain $(\p^2)^2$ on either side of the interface, the dimensions of  $\widehat\p^i$ and $\pa_z\widehat\p^i$ remain classical. Moreover, at this order all fields of the kind $\pa_z^k\wh{\p^i}$ can be converted to descendants of $\widehat\p^i$ and $\pa_z\widehat\p^i$ by means of the tree level equations of motion. Hence, the latters are the only primaries appearing with an OPE coefficient of order one. The interesting fact is that $\wh{\D}_{\p}$ and $\wh{\D}_{\pa\p}$ do not receive loop corrections at all, as we review in subsection \ref{subsec:interboot}. A last comment on the one-loop odd spectrum is in order. The two-point function of $\p^2\p^i$ should obey eq. \eqref{eq:2pointsame} only on the free side, where the operator is a primary. This two-point function contains a tower of defect operators which we might identify with $\widehat{\p^2\p^i}$ and its transverse derivatives. The dimension of $\wh{\p^2\p^i}$ is consistently half-way between $\p^2\p^i$ and its image under RG flow, that is, $\Box \p^i$, and turns out to be marginal at this order. Since we could not devise a mechanism to protect this operator from quantum corrections, we believe this feature will disappear from the spectrum at higher orders. The fact that $\wh{\p^2\p^i}$ is independent from the conformal families of $\wh{\p^i}$ and $\pa^2_z\wh{\p^i}$ is naturally justified by defining the defect fields as the limit of the \emph{free} bulk fields approaching the interface. Notice that this happens automatically in a hard-core regularization, where all integrals are cut-off at a small distance from the interface.

The considerations leading to eq. \eqref{eq:2pointsame} apply in fact to the leading order in conformal perturbation theory of any interface obtained by a nearly marginal bulk perturbation. Indeed, the key point is that the Zamolodchikov norm of the displacement operator equals the square of the coupling at leading order. We turn now to this more general setting in order to discuss the leading order mixing of bulk and defect primaries. On the contrary, notice that eqs. \eqref{eq:across2} and \eqref{eq:across>2} do not generalize trivially, because we used the fact that UV scale dimensions are (nearly) even-integer separated: formulae get a bit more messy in the general case.

\subsection{Leading order mixing of primary operators.}
\label{subsec:mixing}

Consider a conformal field theory in any number of dimensions $d$, whose spectrum includes one\footnote{We consider for simplicity the case of a one parameter RG flow. The general case proceeds along the same lines.} mildly relevant operator $\vp$, that is $\ep=d-\D_\vp$ is a small positive number. The interface constructed by integrating $g\vp$ on one half of the space has an infrared fixed point in which $g=g^*\sim \mc{O}(\ep).$ The two-point functions of operators on the same side of the interface obey the obvious generalization of eq. \eqref{eq:2pointsame}:
\begin{multline}
\braket{O_1(x)O_2(x')} \\
=\frac{1}{(2x_d)^{\D_1}(2x'_d)^{\D_2}}\,\xi^{-(\D_1+\D_2)/2}
\left(\de_{12}+\la_{12\vp}a_{\vp} \fk^{d}(\D_{12},\D=d,\xi)\right)
+\mc{O}(\ep^2).
\label{eq:2pointsamed}
\end{multline}
Here $a_\vp$ is of order $\ep$ and at this order
\beq
a^\textup{IR}=-a^\textup{UV}=g^*\,\frac{S_d}{d},
\label{eq:avp}
\eeq
as dictated by eq. \eqref{eq:scalingOnepoint}. We would like to study the effect of the mixing of bulk primaries on the defect operators. Let us choose a set of UV scalar primaries $O^\textup{UV}_i$ which are degenerate up to terms of order $\ep$. Their defect OPE, restricted to the lowest lying primaries, is encoded in the fusion rule
\beq
O^\textup{UV}_i \sim \m_i{}^j\, \wh{O}_j+\dots
\label{eq:UVfusion}
\eeq
These defect operators are connected by the RG flow to the UV operators themselves, that is there exists a family of renormalized operators $\wh{O}_i(g)$ such that $O^\textup{UV}_i=\wh{O}_i(0)$ and $\wh{O}_i=~\wh{P}_i{}^j\,\wh{O}_j(g^*)$. The matrix $\wh{P}^i{}_j$ depends on the definition of the renormalized operators, that is on the regularization scheme. However, in what follows we shall only need the fact that $\wh{P}^i{}_j$ is orthogonal at order one. Comparing with eq. \eqref{eq:UVfusion} we see that
\beq
\m_i{}^j=\wh{P}^j{}_i+\mc{O}(\ep).
\eeq
The relevant part of the defect OPE decomposition of the correlator $\braket{O^\textup{UV}_iO^\textup{UV}_j}$ is determined by the following asymptotic behavior for large $\xi$:
\beq
\fk^{d}(\D_{12}=0,\D=d,\xi)\ \sim\ \frac{d}{2}\big(\log \xi + \gamma - 
\psi(d/2)\big) +
\mc{O}(\xi^{-1}).
\eeq
Comparing this with the large $\xi$ and small $\ep$ limit of the boundary blocks, we get
\beq
\sum_k \m_i{}^k\mu_{jk} \left(\wh{\D}_k-\frac{\D^\textup{UV}_i+\D^\textup{UV}_j}{2}\right)
=-\frac{d}{2}\,\la^\textup{UV}_{ij\vp}\,a^\textup{UV}_{\vp}.
\label{eq:UVleadingDOPE}
\eeq
Since the quantity in parenthesis is of order $\ep$, we can make the substitution $\m\to \wh{P}$. The latter matrix was defined to be the orthonormal change of basis which diagonalizes the matrix of anomalous dimensions $\wh{\ga}_i{}^j$ of the boundary operators $\wh{O}_j(g)$, so that we get
\beq
\wh{\ga}_{ij}= \D^\textup{UV}_i\de_{ij}- \frac{d}{2}\,\la^\textup{UV}_{ij\vp}\,a^\textup{UV}_{\vp}
= \D^\textup{UV}_i\de_{ij} + \frac{S_d}{2}\,\la^\textup{UV}_{ij\vp}\, g^*.
\label{eq:mixingdefect}
\eeq
One may proceed order by order in the large $\xi$ expansion. The resulting defect spectrum includes in general nearly degenerate scalars with dimension close to $\D+k$, $\D$ being the scale dimension of a bulk primary. A primary of level $k$ of course originates from linear combinations of transverse and parallel derivatives of a UV primary. But when nearly integer separated bulk primaries exist, further mixing is expected to take place.

To complete the analysis, we would like to show that by matching the defect spectrum with the IR bulk primaries, we get back the known mixing matrix between UV and IR operators of the homogeneous theory \cite{Zamolodchikov:1987ti}. We restrict ourselves to the case in which the mixing only involves primary operators. We consider the set of IR primaries $O^\textup{IR}_i$ which are related to the $O^\textup{UV}_i$ through a matrix $P^i{}_j$ whose definition is analogous to the one we gave for $\wh{P}$. The leading part of the defect fusion rule is
\beq
O^\textup{IR}_i \sim \n_i{}^j \wh{O}_j + \dots
\eeq
where we required that the defect spectrum coincides with the one of the UV counterparts. This time we have
\beq
\n_i{}^j = P_i{}^k \wh{P}^j{}_k+\mc{O}(\ep).
\label{eq:IRdopeCoeff}
\eeq
The same steps as before now lead to a relation identical to eq. \eqref{eq:UVleadingDOPE}, up to the substitutions $\m\to \n$ and UV $\to$ IR. The combination of eqs. \eqref{eq:avp}, \eqref{eq:mixingdefect}, \eqref{eq:IRdopeCoeff}  with the statement
\beq
\la^\textup{IR}_{ij\vp}= P_i{}^m P_j{}^n \la^\textup{UV}_{mn\vp} +\mc{O}(\ep),
\eeq
leads to
\beq
\D^\textup{IR}_i \de_{ij} 
= P_i{}^m P_j{}^n \left( \D^\textup{UV}_m \de_{mn}+S_d\,\la^\textup{UV}_{mn\vp}\, g^* \right).
\eeq
Since the matrix $P$ diagonalizes by hypothesis the matrix of bulk anomalous dimensions, we recover the formula
\beq
\ga_{ij} =  \D^\textup{UV}_i \de_{ij}+S_d\,\la^\textup{UV}_{ij\vp}\, g^*.
\label{eq:mixingbulk}
\eeq
Notice that the anomalous part of the defect mixing matrix is one half of the bulk one. 

As a last comment, by means of eq. \eqref{eq:scalingAcross}, we can verify that the pairing of UV and IR primaries matches the matrix $P$ at leading order \cite{Gaiotto:2012}:
\beq
a_{ji}=P_{ij}+\mc{O}(\ep).
\label{eq:matching}
\eeq
Indeed, eq. \eqref{eq:matching} is immediately obtained starting from the equality
\beq
(\D^\textup{IR}_i-\D^\textup{UV}_j) a_{ji}= S_d\, P_i{}^k \la^\textup{UV}_{jk\vp}\,g^*,
\eeq
which is valid at leading order, and using the definition \eqref{eq:mixingbulk} of the mixing matrix.

\subsection{The interface bootstrap.}
\label{subsec:interboot}

In order to single out a solution to the crossing equation which corresponds to our interface, we shall again concentrate on the 3d Ising model, and in particular on the two-point functions involving the lowest lying odd primaries $\p$ and $\si$, on the free and interacting side respectively. The bootstrap constraints involving $\p$ can be in fact completely solved in any number of dimensions by requiring the correlation functions to be annihilated by the Laplace operator. Therefore, we start by collecting some general facts about free bosonic theories in the presence of codimension one conformal defects. Let us first of all consider the two-point function $\braket{\p\p}$. As it is well known, one can prove by applying the equations of motion to the $\p\times \p$ OPE that it contains only twist two operators, and in particular:
\beq
\p\times\p \sim 1+ \p^2 + (\textup{primaries with zero expectation value}).
\eeq
The same method can be applied for establishing that only two primaries appear in the defect OPE of the field (this was first noticed in \cite{Dimofte:2012pd}). Indeed, when the Laplace operator is applied to the r.h.s. of the defect OPE, the parallel derivatives give descendants and we can disregard them. The derivative orthogonal to the defect imposes a constraint on the scale dimension of allowed primaries:
\beq
0 =\square  \p (\bm{x},x^d) \sim \sum_{\wh{O}} (\D_{\wh{O}}-\D_\p)\,(\D_{\wh{O}}-\D_\p-1)\,
\frac{\wh{O}(\bm{x})}{(x^d)^{\D_\p-\D_{\wh{O}}+2}}+\textup{descendants}
\eeq
Hence, there are only two primaries, the limiting value of the field $\wh{\p}$ and of its derivative $\wh{\pa\p}$. These primaries have protected dimensions $\D_{\wh{\p}}=\frac{d}{2}-1$ and $\D_{\wh{\pa\p}}=\frac{d}{2}.$ We see that the most general defect CFT featuring the free theory on half of the space, bounded by any codimension one defect, satisfies the following crossing equation:
\beq
1+\la_{\p\p\p^2}\,a_{\p^2} \fk^d(\D_{\p^2},\xi) = 
\xi^{\D_\p} \left( \m^2_{\p\wh{\p}}\, \fy^d(\D_{\wh{\p}},\xi)
+\m^2_{\p\wh{\pa\p}}\, \fy^d(\D_{\wh{\pa\p}},\xi) \right).
\eeq
All conformal blocks reduce to elementary function:
%\footnote{A little care is needed for the conformal block of $\wh{\p}$ \cite{McAvity:1995zd}.}
\begin{align}
\fy^d\left(\frac{d-2}{2},\xi\right) &=
     \frac{1}{2}\xi^{-\D_\p}\left(1+\bigg(\frac{\xi}{\xi+1}\bigg)^{\D_\p}\right) \\
\fy^d\left(\frac{d}{2},\xi\right) &=
 \frac{2}{d-2}\xi^{-\D_\p}\left(1-\bigg(\frac{\xi}{\xi+1}\bigg)^{\D_\p}\right),
\end{align}
so the crossing equation is equivalent to the following:
\begin{subequations}
\begin{align}
\frac{1}{2}\, \m^2_{\p\wh{\p}} + \frac{2}{d-2}\, \m^2_{\p\wh{\pa\p}}  &=  1, \\
\frac{1}{2}\, \m^2_{\p\wh{\p}} - \frac{2}{d-2}\, \m^2_{\p\wh{\pa\p}}   &= \la_{\p\p\p^2}\,a_{\p^2}.
\end{align}
\label{eqn:freeboot:ppsystem}
\end{subequations}
The solution is parametrized by an angle:
\beq
\m_{\p\wh{\p}} = \sqrt{2}\,\cos \al, \quad
\m_{\p\wh{\pa\p}} = \sqrt{\frac{d-2}{2}}\,\sin \al, \quad
\la_{\p\p\p^2}\,a_{\p^2} = \cos 2\al.
\label{eqn:freeboot:ppalpha}
\eeq
The solution of this particular crossing equation is only a necessary condition for the existence of a full fledged defect CFT, therefore the question arises whether for any value of $\al$ such a theory exists. Vice versa, a given value of $\al$ might be realized in more than one defect CFT, which differ elsewhere. We can restrict $\al$ to take values in the interval $[0,\pi/2]$, since sending the defect fields $\wh{\p}$ and $\wh{\pa\p}$ to minus themselves does not spoil their canonical normalization. At the extrema of this interval one finds Neumann ($\al=0$) and Dirichlet ($\al=\pi/2$) boundary conditions, and at the center ($\al=\pi/4$) the trivial interface between the free theory and itself. The RG interface with the $O(N)$ model with $\p^4$ interaction lies perturbatively near to the no-interface value, in $\ep$-expansion, and fills an interval if $N$ is allowed to take value over the reals.

Since any two-point function involving the field $\p$ has to contain only the same two blocks in the defect channel, one can generalize the previous procedure to any correlator of this kind. The general fusion rule with a primary $O$ with dimension $\D$ is
\begin{multline}
\p \times O_{\D} \sim O_{-}+O_{+}+
(\textup{spinning primaries}), \\
\D^-=\D-\D_\p,  \quad  \D^+=\D+\D_\p.
\label{eq:pope}
\end{multline}
Notice that degenerate primaries may exist with the right dimensions to enter the r.h.s. of eq. \eqref{eq:pope}, as it happens in the $O(N)$ model for $N>1$. Denoting $\la_+=\la_{\p O_{\D} O_+}$ and $\la_-= \la_{\p O_{\D} O_-}$, the solution to the bootstrap equation is
\beq
\m_{\p\wh{\p}}\,\m_{O\wh{\p}} =  
\la_-\,a_- + \la_+\,a_+, \qquad
\frac{4}{d-2}\, \m_{\p \wh{\pa\p}}\, \m_{O \wh{\pa\p}}   =  \la_-\,a_- - \la_+\,a_+.
\label{eqn:freeboot:systemgeneral}
\eeq
This includes the system \eqref{eqn:freeboot:ppsystem}, in particular. The relations \eqref{eqn:freeboot:systemgeneral} also apply when the operator $O$ is a primary on the interacting side of the interface. In this case, the OPE happens in the folded picture, and turns out to be a simple way to choose the solution of the Laplace equation with the appropriate asymptotics. Specifically, no singularities should arise when the operators are placed in mirroring points, and this prompts us to eliminate $O_-$ from the r.h.s. of eq. \eqref{eq:pope}. In other words,
\beq
\p\times O \sim\ :\!\p O\!: \,,
\eeq
and the two-point function is simply
\beq
\braket{\p(x)O(x')} = \frac{a_{\p O}}{(2|x^d|)^{\D_\p}(2x'^d)^{\D}}
 \   {}_2F_1\left(\D_\p,\D,\D,-\xi_\textup{folded}\right) 
 = \frac{a_{\p O}}{(2x'^d)^{\D-\D_\p}(x-x')^{2\D_\p}},
\eeq
where $\xi_\textup{folded}$ is just obtained by replacing $x^d$ with minus itself. The relation \eqref{eqn:freeboot:systemgeneral} reduces to
\beq
a_{\p O}=\m_{\p\wh{\p}}\, \m_{O\wh{\p}}=-\frac{4}{d-2} \m_{\p \widehat{\pa \p}}\, \m_{O\widehat{\pa \p}}.
\label{eqn:freeboot:pOdata}
\eeq
This relation is potentially useful in bootstrapping the interacting side of the interface. Indeed, the defect OPE of every operator which couples with $\p$ contains $\wh{\p}$ and $\wh{\pa\p}$, and the ratio $\m_{O\wh{\p}}/\m_{O\wh{\pa\p}}=-2\,\tan \al/ \sqrt{d-2}$ does not depend on the operator,  and may be used to match solutions for different external primaries. From eq. \eqref{eqn:freeboot:ppsystem}, we see that this ratio among coefficients of the interacting theory is determined by the expectation value of $\p^2$ on the free side. In particular, as we pointed out, this one-point function deviates from zero only at order $\ep^2$ in the case we are interested in. We compute the leading order value in appendix \ref{sec:details} for generic $N$, and find
\beq
\al = \frac{\pi}{4} - \frac{3}{1024 \pi^6}\frac{N+2}{(N+8)^2}\ep^2.
\eeq
In sum, the signature of the RG domain wall in the conformal block decomposition of $\braket{\si\si}$ is the presence of two protected defect operators, with a ratio of OPE coefficients near to the free theory value.

In fact, we found in $3d$ a numerical solution for a (4,4,0) truncation of 
$\braket{\si\si}$ which has the expected features. The defect channel is formed by the two operators $\widehat{\si}$ and  $\widehat{\partial_z\si}$ of protected 
dimensions 
$\frac12$ and $\frac32$ and two unprotected operators $\widehat{O}_3$ and 
$\widehat{O}_4$  of dimensions
$\widehat{\D_3}\sim 3.11$  and $\widehat{\D_4}\sim6.17$. The precise value of these
quantities as well as the estimates of the relative OPE coefficients depend on the 
choice of the bulk spectrum. For the sake of consistency we put in the same bulk spectrum 
obtained in the (4,2,1) solution of the extraordinary transition. The values of
$\D_{\varepsilon''}$ and $\D_{\varepsilon'''}$ depend on the scale dimension
$\hat{\D}$ of a surface operator which acts as a free parameter. Therefore,  
our interface solution also depends on it, though the dependence 
is very mild, as a stable solution requires (see the discussion on the stability of 
the solutions on section \ref{sec:method}). 
Table \ref{tab:3} shows the relevant data of such a solution. Note that the ratio
$ \mu_{\si\widehat{\si}}/\mu_{\si\widehat{\partial_z\si}}$ follows the trend suggested by 
the $\ep$ expansion. 

\begin{table}[htb]
\centering
\begin{tabular}{*{5}{>{$}c<{$}}}
\toprule
\hat{\Delta}&\Delta_{\epsilon''}&\Delta_{\epsilon'''}&\mu^2_{\si\widehat{\si}}
&\mu^2_
{\si\widehat{\partial_z\si}}\\
\midrule
    3.9 & 7.235(6)(3) & 12.736(7)(4) & 1.00612(11)(5) & 0.27138(5)(2) \\
    7.9 & 7.274(10)(2) & 12.843(17)(4) & 1.00644(15)(4) & 0.27123(7)(1) \\
    11.1 & 7.287(11)(4) & 12.892(22)(8) & 1.00657(17)(6) & 0.27117(7)(2) \\
    15.1 & 7.297(11)(2) & 12.932(23)(4) & 1.00668(16)(3) & 0.27112(7)(2) \\
    19.9 & 7.298(11)(1) & 12.948(24)(2) & 1.00667(16)(2) & 0.271127(68)(5) \\
    25.5 & 7.302(11)(2) & 12.968(25)(5) & 1.00672(16)(3) & 0.27110(7)(2) \\
   31.9 & 7.303(11)(3) & 12.980(25)(7) & 1.00674(16)(5) & 0.27110(7)(1) \\
    39.1 & 7.307(12)(4) & 12.995(28)(8) & 1.00679(17)(5) & 0.27108(7)(2) \\
 \end{tabular}
\bigskip
\begin{tabular}{*{5}{>{$}c<{$}}}
\toprule
\hat{\Delta}& \widehat{\Delta}_3& \widehat{\Delta}_4&\mu^2_{\si\widehat{O}_3}
&\mu^2_{\si\widehat{O}_4}\\
\midrule
    3.9 & 3.1190(8)(4) & 6.1816(9)(4) & 0.002555(5)(2) & 0.00002387(4)(2) \\
    7.9 & 3.1151(12)(3) & 6.1757(15)(4) &0.002572(7)(2) & 0.00002408(6)(2) \\
    11.1 & 3.1136(14)(5) & 6.1734(17)(6) &0.002579(7)(3) & 0.00002417(7)(3) \\
    15.1 & 3.1123(14)(3) & 6.1715(17)(3) &0.002584(7)(2) & 0.00002424(7)(2) \\
    19.9 & 3.1121(14)(1) & 6.1710(17)(2) &0.0025850(74)(10) & 0.00002426(7)(1)\\
    25.5 & 3.1115(14)(3) & 6.1701(18)(4) &0.002588(7)(1) & 0.00002429(7)(1) \\
    31.9 & 3.1113(14)(4) & 6.1697(17)(5) & 0.002589(7)(2) & 0.00002431(7)(3) \\
    39.1 & 3.1108(15)(5) & 6.1689(19)(6) & 0.002591(8)(2) & 0.00002433(8)(2) \\
 \end{tabular}
\caption{Data of the (4,4,0) solution of the 3d Ising interface with the free UV theory.
The first column is the free parameter of the solution which is the scale dimension of a 
surface operator contributing to the extraordinary transition discussed in sec. 
\ref{sec:boundary}.
The data are affected by two kinds of errors. The first  parenthesis reflects the 
statistical error of the input data (namely $\D_\si$ 
and $\D_{\varepsilon}$), while the second parenthesis indicates the spread of the solutions.}
\label{tab:3}
\end{table}

Let us make some final remarks. When the bulk OPE coefficients and the scale dimensions are exactly known on one side of an interface, one may extract the one-point functions from the crossing equations involving operators placed on this side. The same data enter various correlators, and the interplay between different solutions to the crossing equations may be used to detect systematics, or to reduce the unknowns. We leave this for future work. For now, we notice that the even spectrum on the free side of our interface is made by an increasing number of degenerate primaries of integer dimension, so it is foreseeable that a reliable truncation would require the inclusion of many bulk primaries. Furthermore, since the parameter $N$ only enters the determinants through the unknown defect spectrum, one expects to find a one-parameter family of solutions.  Studying two-point functions of free even primaries is important in particular if one is interested in the Zamolodchikov norm of the displacement operator. Indeed, two defect primaries exist with dimension $d$, one of which might be identified with the displacement of the folded theory. Given two primaries $O_\textup{L}$ and $O_\textup{R}$ with non-vanishing one-point function, it is not difficult to see that, in order to isolate the displacement, one needs to know $\braket{O_\textup{L}O_\textup{L}}$, $\braket{O_\textup{R}O_\textup{R}}$ and $\braket{O_\textup{L}O_\textup{R}}$. Unfortunately, we have not been able to identify a solution for $\braket{\ep\,\ep}$ which satisfactorily reproduces the domain wall.
\section{Conclusions and outlook.}
\label{sec:conclusions}

In this paper we explored some consequences of crossing symmetry for 
defect CFTs. We focused our study on the cases where the defect is a codimension one hyperplane, i.e. a flat interface or a boundary. In the latter case our main results concern the surface transitions of 3d Ising model. 

The numerical solutions to the bootstrap equations with the method of determinants turn out to be particularly effective in the ordinary transition, where  it suffices to know the scale dimensions of the first few bulk primaries to obtain the dimension of the relevant surface operator of this transition as well as its OPE coefficient. This analysis has been extended to the $O(N)$ models with $N=0,1,2,3$ where a comparison can be  made with the results of a two-loop calculation \cite{Diehl:1998mh}, finding  a perfect agreement (see table \ref{tab:1}).   

In the extraordinary transition the contribution of the boundary channel is dominated by the first two low-lying operators, namely the identity and the displacement, thus we used this fact to extract more information on the  even and odd spectrum  contributing to the bulk channel. We obtained in this way also an accurate determination of the OPE coefficient $\lambda_{\si\si\varepsilon}$ which 
compares well with other estimates based on a recent  Monte Carlo 
calculation \cite{Caselle:2015csa} or on conformal bootstrap \cite{El-Showk:2014dwa}. We also obtained some OPE coefficients of one-point and two-point functions  (see table \ref{tab:2}) which allow to verify the impressive fulfillment of the Ward identities associated with the displacement operator.

The solution corresponding to the special transition contains a free parameter, hence we don't get precise numerical results. This case is still very useful for an accurate cross-check of the consistency of the method of determinants with the linear functional method. Together with the just mentioned Ward identities, this check provides evidence for the fact that the systematic error is rather small when a truncation is stable. In this paper we investigated the stability of the truncations through the sensitivity to the addition of heavier operators. It would be important to establish more rigorous bounds on the systematic error, maybe along the lines of \cite{Pappadopulo:2012jk}. 

The next example of a codimension one defect studied in this paper is an interface between the $O(N)$ model and the free theory. We tackled the problem both in $4-\ep$ and in three dimensions. The weak coupling analysis of the two-point functions was carried out in a way which is trivially adapted to general perturbation interfaces. A preeminent role is played by the displacement operator, whose small Zamolodchikov norm signals the transparency of the interface, in the sense that operators with nearly degenerate dimension are allowed to couple at order one across the interface, while the opposite is true for primaries well separated in the spectrum. This intuition can be made precise in $2d$, where the norm of the displacement coincides up to a normalization with the reflection coefficient defined in \cite{Quella:2006de}.\footnote{In particular, it is not difficult to prove unitarity bounds for reflection and transmission in function of the central charges, just by diagonalizing the defect spectrum.} It is certainly interesting to look for a similar interpretation of the displacement in higher dimensions, possibly in relation to the correlators of polarized stress-tensors. However, it is worth emphasizing that while in $2d$ the reflection coefficient of a boundary is unity, in dimensions greater than two the norm of the displacement depends on the boundary conditions. The results of the perturbative analysis also confirm that this kind of interfaces encode information about the RG flow that links the theories on the two sides: specifically, the coupling of UV and IR primaries reproduces the leading order mixing of operators, as does the one-dimensional domain wall constructed non-perturbatively in \cite{Gaiotto:2012}. On the numerical side, we found a solution to the crossing equation consistent with the features of the two-point function of $\si$ in three dimensions. The analysis can be extended in various directions. It would be interesting go to second order in perturbation theory \cite{Gaberdiel:2008fn}, or to study the setting at large $N$, and see whether the displacement operator still provides important simplifications. We already pointed out that it is viable to bootstrap correlators on the free side, and it would be important in particular to give a prediction for the norm of the displacement in $3d$, to compare it with the estimates for the boundary transitions. We would also like to emphasize that the interface can be realized on the lattice, for instance as a Gaussian model with the addition of a quartic potential on one-half of the lattice.

As we mentioned in the introduction, a complete description of the CFT data cannot be reached, even in principle, only through the study of bulk two-point functions. Four-~point functions of defect operators should be studied, and in this case both the method of determinants and the linear functional might be employed. Along the same lines, in both the boundary and the interface setups one may study the crossing constraints coming from correlators of the kind $\braket{O_1 O_2 \widehat O}$, or two-point functions of tensors. The necessary tools for the latters were developed in \cite{Liendo:2012hy}. It is of course viable to use the method of determinants for the study of generic defects, and in particular it would be nice to complement the bootstrap analysis carried out in \cite{Gaiotto:2013nva} for the twist line in the Ising model.

\section*{Acknowledgements}
We thank Leonardo Rastelli for pointing out  an error in the first version of this paper.
Preliminary results of this paper were first presented at the workshop 
``Back to the Bootstrap IV'' at Porto University, June 30-July 11, 2014. FG would like to 
thank the organizers and the participants for the stimulating atmosphere; he also thanks John 
Cardy and Slava Rychkov for fruitful discussions. PL would like to
thank Leonardo Rastelli and Balt van Rees for discussions during the
early stages of this work. MM would like to thank Michele Caselle,
Dalimil Mazac, Enrico Trincherini and Ettore Vicari for useful
discussions, and especially Davide Gaiotto, for suggesting the study of the RG interface and for many illuminating
discussions. He also thanks Perimeter Institute for Theoretical
Physics for the hospitality during the preparation of this paper. PL
is supported by SFB 647 ``Raum-Zeit-Materie. Analytische und
Geometrische Strukturen''. AR is supported by the Leverhulme Trust (grant RPG-2014-118) and STFC (grant ST/L000350/1). Research at Perimeter Institute is
supported by the Government of Canada through Industry Canada and by
the Province of Ontario through the Ministry of Research and
Innovation. 

\appendix

\section{RG domain wall: details on the $\ep$-expansion.}
\label{sec:details}

\subsection{One loop computations.}

Two regularization procedures have been preferred in the literature, in dealing with the $\p^4$ model in the presence of a defect of co-dimension one. Dimensional regularization has been especially used for the systematic renormalization of the Lagrangian and for extracting the critical exponents \cite{Diehl:1981,Diehl:1983,Diehl2:1983}. 
More recently, fully real space computations were carried out in \cite{McAvity:1993ue, McAvity:1995zd}, with a short distance cutoff. Both series of works were concerned with the $\p^4$ theory in the presence of a plain boundary. We follow the latter technique.

\begin{figure}[b]
\centering
\includegraphics[height=0.2\textheight,width=0.47\columnwidth]{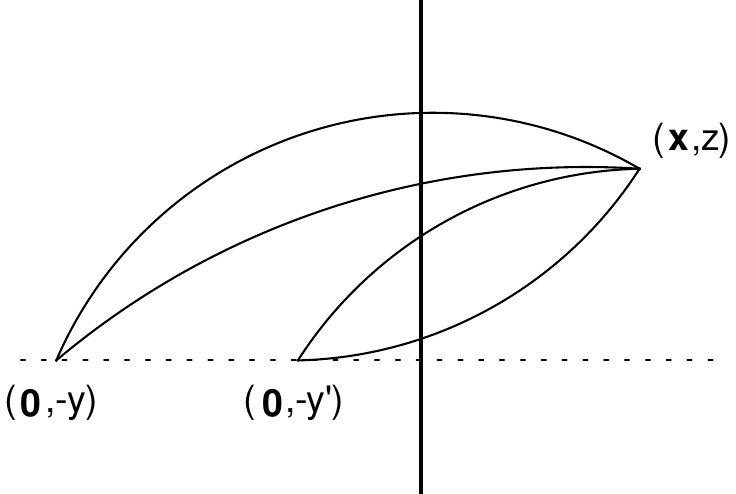}
\includegraphics[height=0.2\textheight,width=0.47\columnwidth]{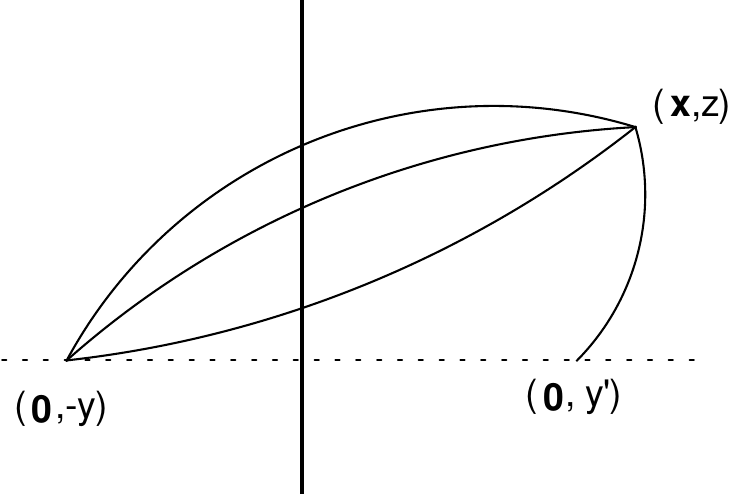}
\caption{One loop contributions to $\braket{\p^2(x)\p^2(x')}$ and to $\braket{\p^2\p^i(x)\p^j(x')}.$ The free side is the left one, and $y,y',z>0.$}
\label{fig:oneloop}
\end{figure}

We start by checking eq. \eqref{eq:2pointsame} through the two-point function $\braket{\p^2\p^2}$ on the free side of the domain wall. At one loop, the only diagram contributing is shown in fig. \ref{fig:oneloop}. 
Since the correlator depends only on one cross-ratio, it is sufficient \cite{McAvity:1993ue} to compute the two-point function in the collinear geometry of fig. \ref{fig:oneloop}, for which
\beq
\xi \to \frac{(y-y')^2}{4y y'}.
\eeq
The corresponding integral is
\begin{multline}
\braket{\p^2(x)\p^2(x')}_\textup{one-loop} \\
=-\frac{1}{3}N(N+2)g^* \int_0^\infty\! dz\,\int\! d^{d-1}\bm{x}\,
\frac{1}{\left\{\left(\bm{x}^2+(z+y)^2\right)\left(\bm{x}^2+(z+y')^2\right)\right\}^{d-2}}.
\label{eqn:interf:p2p2oneloop_int}
\end{multline}
Notice that we chose $y,\,y'>0$. The integral does not diverge in the UV.
This is expected, since the coupling constant renormalizes at $O(\ep^2)$, and the lowest lying interface operator that might be needed as a counterterm is $\phi^4$, which however - barring mixing which appears at higher orders - equals the displacement operator and is therefore irrelevant. Since the fixed point coupling constant $g^*$ is of order $\ep$, we can plug $d=4$ in the integral to obtain the leading order correction, which is easily computed. The result is
\beq
\braket{\p^2(x)\p^2(x')}_\textup{one-loop}
= \frac{N(N+2)}{3}\,g^*  \frac{\pi^2}{(y-y')^4}\left(\frac{\xi}{\xi+1}-\log (1+\xi)\right)+O(\ep^2).
\eeq
Plugging into this expression the fixed point value for $g^*$ \eqref{eq:gstar} and adding the tree level contribution, one obtains the correlator at first order in $\ep$-expansion:
\beq
\braket{\p^2(x)\p^2(x')}=\frac{2 N}{s^{2(d-2)}}\left\{1+
\frac{1}{2}\frac{N+2}{N+8}\,\ep\,\left(\frac{\xi}{\xi+1}-\log (1+\xi)\right)\right\}.
\label{eq:p2p2oneloop_res}
\eeq
Notice that the one point function of $\p^2$ is $O(\ep^2)$, therefore this is the full correlator - not just the connected part - at order $\ep$. Comparing the result with the form of the conformal block of $\p^4$, evaluated in $d=4$ at this order:
\beq
\fk^{d=4} (\D_{\p^4};\xi)=-2 \left(\frac{\xi}{\xi+1}-\log (1+\xi)\right),
\eeq
and using that in free theory
\beq
\la_{\p^2\p^2\p^4} = \sqrt{\frac{2 (N+2)}{N}},
\eeq
we see agreement with the general result \eqref{eq:2pointsame} and with the one-point function in eq. \eqref{eq:aUVaIR}.

Let us compare also the general formula \eqref{eq:across2} with an explicit one-loop example. We focus on the correlator between the field $\p^i$ on the interacting side and the free primary $\p^2\p^i$. The one-loop contribution is encoded in the diagram on the right in fig. \ref{fig:oneloop}, which is UV finite. Including the combinatorics, the result is
\beq
\braket{\frac{\p^2\p^i(x)}{\sqrt{2(N+2)}}\ \p^j(x')} = \frac{\de^{ij}}{(-2y)^3(2y')}\,\xi^{-2}\, \frac{\sqrt{N+2}}{2\sqrt{2}(N+8)}\, \ep\, (\xi-1).
\label{eq:p3p}
\eeq
It is easy to compute the tree level three-point function needed to fix $\m_{\textup{L\,R}D}$, and see that eq. \eqref{eq:p3p} matches eq. \eqref{eq:across2}.

\begin{figure}[b]
\centering
\includegraphics[height=0.2\textheight,width=0.55\columnwidth]{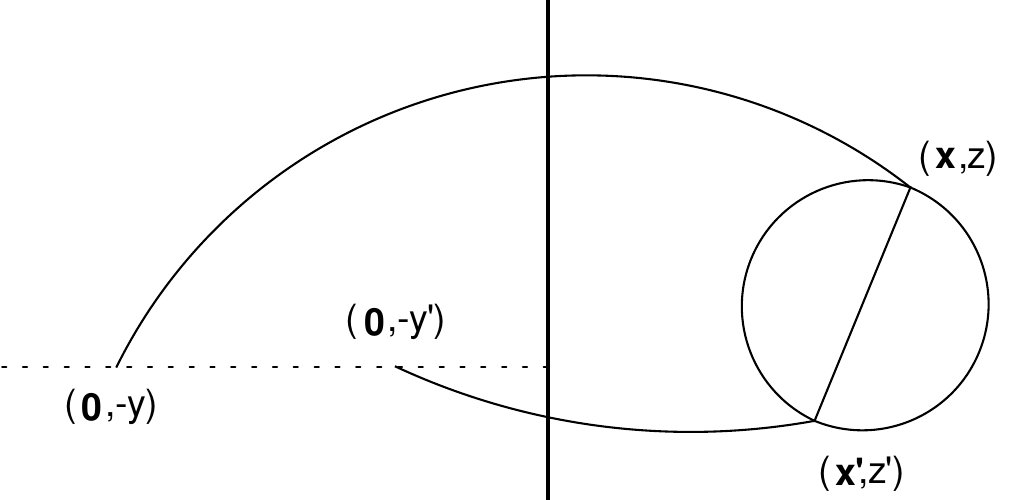}
\caption{Two loops contribution to $\braket{\p^i(x)\p^j(x')}.$ Again, $y,y',z,z'>0.$}
\label{fig:interf:ppfree}
\end{figure}

Next, we compute the first non-trivial contribution to the two-point function of $\p^i$ on the free side, which departs from its free theory value at order $\ep^2$. The only diagram contributing is the sunset (fig. \ref{fig:interf:ppfree}). 
As explained in subsection \ref{subsec:interboot}, we actually only need to know $a_{\p^2}$, which amounts to colliding the two external operators in the diagram. The computation only slightly simplifies at this order, but the statement is valid at any loop (and of course, for any interface involving the free theory). The bulk conformal block of the operator $\p^2$ of dimension $\D_{\p^2}=d-2$ is
\beq
\fk(\D_{\p^2};\xi)=\left(\frac{\xi}{1+\xi}\right)^{d-2}.
\eeq
Therefore we find
\beq
\braket{\p^i(x)\p^j(x')} = \frac{\de^{ij}}{s^{d-2}}\left(1 + \la_{\p\p\p^2} a_{\p^2}\, \left(\frac{\xi}{\xi+1}\right)^{d-2} \right).
\label{eqn:interf:ppnonpert}
\eeq
The integral to be evaluated is the following:
\begin{multline}
I=\int_0^{\infty}\!\! dz\,\int_0^{\infty}\!\! dz'\,\int\! d^3\bm{x}\int\! d^3\bm{x'}
\frac{1}{\big(\bm{x}^2+(y+z)^2\big)\big(\bm{x'}^2+(y+z')^2\big)} \\
\times\frac{1}{\big((\bm{x}-\bm{x'})^2+(z-z')^2\big)^3}.
\end{multline}
Along the computation, which is straightforward, we encounter two divergences. A bulk divergence requires a mass counterterm, and a second divergence arises when the interaction vertices hit the interface. This is compensated by integrating $\widehat\p^2$ along the interface. Relevant operators are required because our cut-off breaks scale invariance. Their renormalized couplings, however, must be fine-tuned in order to reach the critical point. Hence, requiring scale invariance of the one-point function is sufficient to fix the subtraction unambiguously. After renormalization, one finds
\beq
I= \frac{3\pi^4}{16}\frac{1}{y^2}.
\eeq
Taking the combinatorics into account, the expectation value at leading order is
\beq
\braket{\frac{\p^2(-y)}{\sqrt{2N}}}\equiv \frac{a_{\p^2}}{(2y)^2} = \frac{3}{512 \pi^6}\sqrt{\frac{N}{2}}\frac{N+2}{(N+8)^2}\ep^2\frac{1}{(2y)^2}.
\eeq
Substituting back in \eqref{eqn:interf:ppnonpert}, and using
\beq
\la_{\p\p\p^2}=\sqrt{\frac{2}{N}},
\eeq
we find at this order
\beq
\braket{\p^i(x)\p^j(x')} = \frac{\de^{ij}}{s^{d-2}}\left(1 +\frac{3}{512 \pi^6}\frac{N+2}{(N+8)^2}\ep^2 \, \left(\frac{\xi}{\xi+1}\right)^2 \right).
\label{eqn:interfep:pppert}
\eeq
One can now extract some CFT data. By using the relations \eqref{eqn:freeboot:ppsystem} one finds the defect OPE coefficients
\beq
\m_{\p\,\wh{\p}}= 1+\frac{3}{1024 \pi^6}\frac{N+2}{(N+8)^2}\ep^2  ,  
\quad  
 \m_{\p\,\wh{\pa \p}}= \sqrt{\frac{d-2}{4}}\left( 1- \frac{3}{1024 \pi^6}\frac{N+2}{(N+8)^2}\ep^2 \right).
\eeq
We also obtain a piece of information about the defect OPE of any primary on the interacting side which couples with $\p^i$, through the equalities \eqref{eqn:freeboot:pOdata}:
\beq
\frac{\m_{O\,\wh{\p}}}{\m_{O\,\wh{\pa\p}}}
=-\frac{4}{d-2}\frac{\m_{\p\,\wh{\pa\p}}}{\m_{\p\,\wh{\p}}}
=-\frac{2}{\sqrt{d-2}}\left(1-\frac{3}{512\pi^6}\frac{N+2}{(N+8)^2}\ep^2 \right).
\label{eq:DopeRatio}
\eeq
We use this result in subsection \ref{subsec:interboot} as a check of the solutions to the approximate crossing equation for $\braket{\si\si}$.

\subsection{Two-point functions across the interface.}

We give some details on the formulae \eqref{eq:acrossdeg}, \eqref{eq:across2} and \eqref{eq:across>2}. Let us call $x^d=\yi$ the position of the interface. We choose again the collinear geometry for the two operators and we place one on either side of the interface, at the points $x=(\bm{x},\yl<\yi)$ and $x'=(\bm{x},\yr>\yi)$. After plugging the free theory three-point function in eq. \eqref{eq:2pointWard}, we shall find the two-point function by solving the following equation:
\begin{multline}
\frac{\m_{\textup{L\,R}\,D}}{(\yr-\yl)^{\D_\textup{L}+\D_\textup{R}-4}}\int\!\! d^{3}\bm{z}\, 
\bigr(\bm{z}^2+(\yl-\yi)^2\bigr)^{-\frac{4+\D_{\textup{LR}}}{2}}
\bigr(\bm{z}^2+(\yr-\yi)^2\bigr)^{-\frac{4-\D_{\textup{LR}}}{2}} \\
=  \frac{d}{d\yi} \braket{O_\textup{L}(x)O_\textup{R}(x')}.
\label{eq:wardoneloop}
\end{multline}
First of all, we briefly comment on \eqref{eq:acrossdeg}, that is, on the case $\D_\textup{L\,R}=\mc{O}(\ep)$. Since $\m_{\textup{L\,R}}$ is also at least of order $\ep$, we can plug $\D_\textup{L}=\D_\textup{R}$ in \eqref{eq:wardoneloop}. The integrals are easily evaluated and we get
\beq
\braket{O_\textup{L}(x)O_\textup{R}(x')}=
-\frac{\pi^2\m_{\textup{L\,R}\,D}}{\abs{\yl-\yr}^{2\D_\textup{L}}} 
\log \left|\frac{\yr-\yi}{\yl-\yi}\right|+c(\yr,\yl).
\eeq
The constant of integration $c(\yr,\yl)$ does not depend on the position of the interface. One way to fix it is to require that when the interface stands half-way between the points the correlator takes the form \eqref{eq:twopmirror}:
\beq
c(\yr,\yl)=\frac{a_\textup{L\,R}}{\abs{\yl-\yr}^{\D_\textup{L}+\D_\textup{R}}}.
\eeq
By asking for conformal invariance of this result, one gets back at first order the scaling relation \eqref{eq:scalingAcross}. Eq. \eqref{eq:acrossdeg} is then obtained by reconstructing the correlator for generic choice of the two points through conformal invariance.

Let us now tackle the case of external dimensions differing at order one. The integration in the translational invariant directions is easily recast as the Euler representation of a hypergeometric function:
\begin{multline}
\int\!\! d^{3}\bm{z}\, 
\bigr(\bm{z}^2+(\yl-\yi)^2\bigr)^{-\frac{4+\D_{\textup{LR}}}{2}}
\bigr(\bm{z}^2+(\yr-\yi)^2\bigr)^{-\frac{4-\D_{\textup{LR}}}{2}} \\
=
\frac{\pi^2}{8}\, \abs{\yl-\yi}^{-1-\D_\textup{LR}} \abs{\yr-\yi}^{-4+\D_\textup{LR}}\,
{}_2 F_1\! \left(\frac{3}{2},\,2-\frac{\D_\textup{LR}}{2};\,4;\,1-\Bigr(\frac{\yl-\yi}{\yr-\yi}\Bigr)^2 \right)
\label{eq:intparallel}
\end{multline}
Internal and spacetime symmetries allow to restrict ourselves to the case $\D_\textup{L\,R}=2 k,$ for integer $k$, at this order. Furthermore, there is a clear symmetry for the exchange L $\leftrightarrow$ R, so we only  consider the case $k>0$. Since for $k>1$ the hypergeometric function is a polynomial, we treat separately the case $k=1$. Eq. \eqref{eq:across2} is obtained integrating the position of the interface and again fixing the integration constant in accordance with conformal invariance. When $k=2,3,\dots$ one can write eq, \eqref{eq:wardoneloop} as
\begin{multline}
\braket{O_\textup{L}(x)O_\textup{R}(x')} =\ \frac{\m_{\textup{L\,R}\,D}}{(\yr-\yl)^{\D_\textup{L}+\D_\textup{R}}}\ \times \\
 \int_{(\yl+\yr)/2}^{y_0} d\yi\,
 \frac{3\pi^{3/2}\Gamma \left(k-\frac{1}{2}\right) }{2 \Gamma (k+2)}\frac{(\yr-\yl)^4}{(\yi-\yl)^5}
 \ 
   _2F_1\left(\frac{5}{2},2-k;\frac{3}{2}-k;\frac{(\yr-\yi)^2}{(\yl-\yi)^2}\right) \\
 +
 \frac{a_\textup{L\,R}}{(\yr-\yl)^{\D_\textup{L}+\D_\textup{R}}}.
\label{eq:intermediate>2}
\end{multline}
One can exploit the fact that the hypergeometric function is a polynomial and integrate addend by addend the second line of \eqref{eq:intermediate>2}. In particular, we can choose to put the interface in $y_0=0$. Some simplifications occur because of the following observation. As already pointed out, the value of $a_\textup{L\,R}$ is fixed by the requirement of conformal invariance. On the other hand, any constant piece in the integration has the only effect of shifting $a_\textup{L\,R}$. Therefore, we disregard such pieces, and fix the constant in the end. All together, introducing the scale invariant variable $r=\yl/\yr$, we find
\begin{multline}
\braket{O_\textup{L}(x)O_\textup{R}(x')} =\ \frac{\m_{\textup{L\,R}\,D}}{(\yr-\yl)^{\D_\textup{L}+\D_\textup{R}}}\, \frac{(-1)^k \pi^{5/2}}{(k-1)k^2\Gamma(k+2)\Gamma(-k-1/2)} \,\frac{(r-1)^2}{4r} \times \\
\left\{\left(2 k (r^2- r+1)+(r+1)^2\right) \,
   _2F_1\left(\frac{1}{2},-k;-k-\frac{1}{2};\frac{1}{r^2}\right) \right.\\
\left.-\left(2k(r^2-r)+(1+r)^2\right) \,
   _2F_1\left(\frac{3}{2},-k;-k-\frac{1}{2};\frac{1}{r^2}\right) \right\} \\
+ \frac{\tilde{a}}{(\yr-\yl)^{\D_\textup{L}+\D_\textup{R}}}.
\label{eq:2intermediate>2}
\end{multline}
We only need to enforce invariance under inversions, which amounts to sending $\yr\to 1/\yr$ and $\yl\to 1/\yl$. With the help of standard hypergeometric identities one can check that the first three lines in \eqref{eq:2intermediate>2} are invariant, therefore
\beq
\tilde{a}=0. 
\eeq
Alternatively, one may simply verify that with this choice the relation \eqref{eq:scalingAcross} is fulfilled. The result is not yet explicitly a function of the cross-ratio. The final form eq. \eqref{eq:across>2} can be obtained at the price of some more massage.

\bibliographystyle{JHEP}
%\bibliography{bibliography}{}
\providecommand{\href}[2]{#2}\begingroup\raggedright\endgroup

\end{document}